\pdfoutput=1
\documentclass{vldb}
\usepackage{graphicx}
\usepackage{balance}  
\usepackage{xcolor}
\usepackage{algorithmic}
\usepackage{defpaper2e}
\usepackage{amsmath}
\usepackage{url}
\usepackage{multirow}
\usepackage[boxruled, linesnumbered] {algorithm2e}
\usepackage{parskip}
\usepackage{subcaption}

\newcommand{\Boxiang}[1]{{{\textcolor{black}{\textbf{Boxiang:}}}{\textcolor{blue}{\textbf{#1}}}}}
\newcommand{\Wendy}[1]{{{\textcolor{black}{\textbf{Wendy:}}}{\textcolor{red}{\textbf{#1}}}}}
\newcommand{\Yanying}[1]{{{\textcolor{black}{\textbf{Yanying:}}}{\textcolor{green}{\textbf{#1}}}}}

\newcommand{\Set}[1]{\mathbb{#1}}

\newcommand{\system}{{\sc DANCE}\xspace}

\vldbTitle{Cost-efficient Data Acquisition on
Online Data Marketplaces for Correlation Analysis}
\vldbAuthors{Yanying Li, Haipei Sun, Boxiang Dong, Wendy Hui Wang}
\vldbDOI{https://doi.org/TBD}
\vldbVolume{12}
\vldbNumber{xxx}
\vldbYear{2019}

\begin{document}


\title{Cost-efficient Data Acquisition on
Online Data Marketplaces for Correlation Analysis}



%
%
%
%

\numberofauthors{3} 
\author{
%
%
\alignauthor
Yanying Li, Haipei Sun\\
       \affaddr{Stevens Institute of Technology}\\
       \affaddr{1 Castle Point Terrace}\\
       \affaddr{Hoboken, New Jersey 07307}\\
       \email{yli158,hsun15@stevens.edu}
\alignauthor
Boxiang Dong\\
       \affaddr{Montclair State University}\\
       \affaddr{1 Normal Ave}\\
       \affaddr{Montclair, New Jersey 07043}\\
       \email{dongb@montclair.edu}
\alignauthor
Hui (Wendy) Wang\\
       \affaddr{Stevens Institute of Technology}\\
       \affaddr{1 Castle Point Terrace}\\
       \affaddr{Hoboken, New Jersey 07307}\\
       \email{Hui.Wang@stevens.edu}
}


\maketitle

\begin{abstract}
Incentivized by the enormous economic profits, the data marketplace platform has been proliferated recently. In this paper, we consider the data marketplace setting where a data shopper would like to buy data instances from the data marketplace for correlation analysis of certain attributes. We assume that the data in the marketplace is dirty and not free. The goal is to find the data instances from a large number of datasets in the marketplace whose join result not only is of high-quality and rich join informativeness, but also delivers the best correlation between the requested attributes.   
To achieve this goal, we design \system, a middleware that provides the desired data acquisition service. 
\system consists of two phases: (1) In the {\em off-line} phase, it constructs a two-layer join graph from samples. The join graph consists of the information of the datasets in the marketplace at both schema and instance levels; (2) In the {\em online} phase, it searches for the data instances that satisfy the constraints of data quality, budget, and join informativeness, while maximize the correlation of source and target attribute sets. We prove that the complexity of the search problem is NP-hard, and design a heuristic algorithm based on Markov chain Monte Carlo (MCMC). Experiment results on two benchmark datasets demonstrate the efficiency and effectiveness of our heuristic data acquisition algorithm.
\end{abstract}

\vspace{-0.05in}
\section{Introduction}
\label{sc:intro}

With the explosion in the profits from analyzing the data, data has been recognized as a valuable commodity. 
Recent research \cite{IDC} predicts that the sales of big data and analytics will reach \$187 billion by 2019.
Incentivized by the enormous economic profits, the {\em data marketplace} model was proposed recently \cite{balazinska2011data}. 
In this model, data is considered as asset for purchase and sale. 
Many established companies (e.g. Twitter and Facebook), sell rich, structured (relational) datasets. But such datasets are often prohibitively expensive. Data marketplaces in the cloud enable to sell the (cheaper) data by providing Web platforms for buying and selling data; examples include Microsoft Azure Marketplace \cite{azure} and BDEX \cite{bdex}.



\begin{table*}
\begin{center}
\begin{subfigure}[b]{0.25\textwidth}
\begin{tabular}{|c|c|c|} \hline
Age & Zipcode & Population\\\hline
[35, 40] & 10003 & 7,000\\\hline
[20, 25] & 01002 & 3,500 \\\hline
[55, 60] & 07003 & 1,200\\\hline
[35, 40] & 07003 & 5,800\\\hline
[35, 40] & 07304 & 2,000\\\hline
\end{tabular}
(a) $D_S$: {\em Source} instance 
owned by data shopper Adam 
\end{subfigure}
~
\begin{subfigure}[b]{0.7\textwidth}
\begin{tabular}{c}
\begin{tabular}{cc}
\begin{tabular}{|c|c|c} \cline{1-2}
Zipcode & State & \\ \cline{1-2}
07003 & NJ & correct\\ \cline{1-2}
07304 & NJ & correct\\ \cline{1-2}
10001 & NY & correct\\ \cline{1-2}
10001 & NJ & wrong \\ \cline{1-2}
\end{tabular}
&
\begin{tabular}{|c|c|c|} \hline
State & Disease & \# of cases\\\hline
MA & Flu & 300 \\\hline
NJ & Flu & 400 \\\hline
Florida & Lyme disease & 130 \\\hline
California & Lyme disease & 40 \\\hline
NJ & Lyme disease & 200 \\\hline
\end{tabular}
\\
$D_1$: Zipcode table 
&
$D_2$: Data and statistics of diseases by state
\\
(FD: Zipcode $\rightarrow$ State)
&
\end{tabular}
\\
\begin{tabular}{cc}
\begin{tabular}{|c|c|c|c|} \hline
Gender & Race & Disease & \# of cases \\\hline
M & White & Flu & 200\\\hline
F & Asian & AIDS & 30 \\\hline
M & White & Diabetes &4,000\\\hline
M & Hispanic & Flu & 140 \\\hline
\end{tabular}
&
\begin{tabular}{|c|c|c|c|} \hline
Age & Gender & Race & Population \\\hline
[35,40] & M & White & 400,000\\\hline
[20,25] & F & Asian & 100,000\\\hline
[20,25] & M & White & 300,000\\\hline
[40,45] & M & Hispanic&50,000 \\\hline
\end{tabular}
\\
$D_3$: Data and statistics of diseases 
&
$D_4$: Census dataset of New Jersey (NJ) State
\\
of New Jersey (NJ) State by gender 
&
\end{tabular}
\\
\begin{tabular}{|c|c|c|c|} \hline
Age & Address & Insurance & Disease  \\\hline
[35, 40] & 10 North St. & UnitedHealthCare & Flu \\\hline
[20, 25] & 5 Main St. & MedLife & HIV \\\hline
[35, 40] & 25 South St. & UnitedHealthCare & Flu \\\hline
\end{tabular}
\\
$D_5$: Insurance \& disease data instance
\\
(b) Relevant instances on data marketplace
\end{tabular}
\end{subfigure}
\caption{\label{table:1} An example of data acquisition}
\end{center}
\vspace{-0.1in}
\end{table*}

In this paper, we consider the data shopper who has a particular  type of data analytics needs as {\em correlation analysis of some certain attributes}. He may own a subset of these attributes already. He intends to purchase more attributes from the data marketplace to perform the correlation analysis. 
However, many existing cloud-based data marketplaces (e.g., Microsoft Azure Marketplace \cite{azure} and Google Big Query service \cite{gwarehouse}) do not support efficient exploration of datasets of interest. The data shopper has to identify the relevant data instances through a non-trivial data search process. First, he identifies which datasets are potentially useful for correlation analysis based on the (brief) description of the datasets provided by the data marketplaces, which normally stays at schema level. There may exist multiple datasets that appear relevant. This leads to multiple purchase options. Based on these options, second, the shopper has to decide which purchase option is the best (i.e., it returns the maximum correlation of the requested attributes) under his constraints (e.g., data purchase budget). 
Let us consider the following example for more details. 
 
\nop{Consider a toy example of the data marketplace that consists of  five data sets shown in Table \ref{table:1}. These five data sets are collected from different sources. $D_1$ - $D_4$ are public datasets\footnote{$D_1$ is collected from USPS\footnote{https://tools.usps.com/go/ZipLookup.}, $D_2$ and $D_3$ are collected from US Centers for Disease Control and Prevention (CDC)\footnote{Centers for Disease Control and Prevention FastStats: https://www.cdc.gov/nchs/fastats/default.htm.}, and $D_4$ is collected from US Census Bureau\footnote{United States Census Bureau Population statistics:https://www.census.gov/topics/population.html.}.}, while $D_5$ is owned by a private party. Apparently, these data sets can be joined in different ways. 
The number of possible join paths is exponential to the number of data sets in the data marketplaces. Given such complexity, it is overwhelming to find the data sets in the large-scale data marketplace that best serve the interests of data shoppers. Therefore, it raises vital needs for techniques and tools that support efficient data acquisition (i.e., search for the best data to purchase) on the data marketplace. 
}
\begin{example}
\label{exp:1}
 Consider the data shopper Adam, a data research scientist, who has a hypothesis regarding the correlation between age groups and diseases in New Jersey (NJ),  USA. Adam only has the information of {\em age}, {\em zipcode}, and {\em population} in his own dataset $D_S$ (Table \ref{table:1} (a)). To test his hypothesis, Adam plans to purchase additional data from the marketplace. Assume that he identifies five instances $D_1$ - $D_5$ (Table \ref{table:1} (b)) in the data marketplace as relevant. There are several data purchase options. Five of these options are listed below: 
\begin{itemize}
\item Option 1: Purchase $D_1$ and $D_2$;
\item Option 2: Purchase three attributes (Gender, Disease, \# of cases) of $D_3$ and (Age, Gender, Population) of $D_4$;
\item Option 3: Purchase three attributes (Race, Disease, \# of cases) of $D_3$ and (Age, Race, Population) of $D_4$;
\item Option 4: Purchase $D_5$; 
\item Option 5: Purchase $D_1$, $D_2$, $D_3$, and $D_4$. 
\end{itemize}
\end{example}

For each option, the correlation analysis is performed on the join result of the purchased datasets and the data shopper's local data (optional). In general, the number of possible join paths is exponential to the number of datasets in the data marketplaces. The brute-force method that enumerates all join paths to find the one that delivers the best correlation between a set of attributes is prohibitively expensive. 
Besides the scalability issue, there are several important issues for data purchase on the data marketplace. Next, we discuss them briefly. 

\noindent{\bf Meaningful joins.}  Different joins return different amounts of useful information. For example, consider Option 1 and 2 in Example \ref{exp:1}. Both Option 1 and 2 enable to generate the association between age and disease, but the join result in Option 1 is grouped by zipcode, while that of Option 2 is grouped by gender. Both join results are considered as meaningful, depending on how the shopper will test his hypothesis on the join results. On the other hand, the join result of Option 4 also includes the attributes {\em Age} and {\em Disease}. However, the join result is meaningless,  as it associates the aggregation data with individual records. 

 \noindent{\bf Data quality.} The data on the data marketplaces tends to be dirty (e.g., missing and inconsistent data values) \cite{hague2013market}. In this paper, we consider {\em data consistency} as the main data quality issue. Intuitively, the data quality (in terms of its consistency) is measured as the percentage of data content that is consistent with a set of specific {\em integrity constraints} which take the form of functional dependency (FD) \cite{fan2008conditional,chiang2008discovering}. For example, $D_1$ in Table \ref{table:1} (b) has a FD: {\em Zipcode} $\rightarrow$ {\em State} (i.e., all the same Zipcode values must be associated with the same State value). However, the last record is inconsistent with the FD. Naturally, the  shopper prefers to purchase the data with the highest quality (i.e., lowest inconsistency). Intuitively, there exists the trade-off between data quality and the performance of correlation analysis - purchasing more data may bring more insights for correlation analysis, but it may lead to more errors. A naive solution is to clean the data in the marketplace first; the purchase requests are processed on the cleaned data. However, we will show that data join indeed impacts the data quality significantly. The join of high-quality (low-quality, resp.) data instances can become low-quality (high-quality, resp.) in terms of data inconsistency. Therefore, data quality issue has to be dealt with online during data acquisition. 

\noindent{\bf Data price and budget-constrained purchase.} Data on the marketplaces are not free. We assume the data shopper is equipped with a limited budget for data purchase. Therefore, we follow the query-based pricing model \cite{balazinska2011data} for the data marketplaces: a data shopper submits SQL queries to the data marketplaces and pays for the query results. This pricing model is also used by some cloud-based data marketplaces, e.g., Google Big Query service \cite{gwarehouse}. This pricing model supports the data shopper to purchase the necessary data pieces (e.g., purchase three attributes \{Gender, Disease, \# of cases\} of $D_3$) instead of the whole dataset. 

The issues of data scalability, data quality, join informativeness and the budget constraints make the data purchase from the data marketplaces extremely challenging for the data shoppers, especially for those common users who do not have much experience and expertise in data management. 
The existing works on data exploration \cite{DBLP:journals/pvldb/YangPS11a,qian2012sample,procopiuc2008database} mainly focus on finding the best join paths among multiple datasets in terms of join size and/or informativeness. However, these solutions cannot be easily adapted to the setting of data marketplaces, due to the following reasons. First, it is too expensive to construct the schema graph \cite{qian2012sample} for all the data on the marketplace due to its large scale. Second, the search criteria for join path is different. While the existing works on data exploration try to find the join path of the best informativeness, our work aims to find the best join path that maximizes the correlation of a set of specific attributes, with the presence of the constraints such as data quality and budgets. 

\noindent{\bf Contributions.} 
 We design a middleware service named \system, a \underline{D}ata \underline{A}cquisition framework on o\underline{N}line data market for \underline{C}orr\underline{E}lation analysis, that provides cost-efficient data acquisition service for the budget-conscious search of the high-quality data on the data marketplaces that maximizes the correlation of certain attributes. \system consists of two phases: (1) during the {\em off-line} phase, it constructs a two-layer join graph of the datasets on the data marketplace. The join graph contains the information of these datasets at both schema and instance level. To deal with the data scalability issue of the marketplaces, \system collects samples from the data marketplaces, and constructs the join graph from the samples; (2) during the {\em online} phase, it accepts the data shoppers' correlation request as well as the constraints (budgets, data quality, and join informativeness). By leveraging the join graph, it estimates the correlation of a certain set of attributes, as well as the join informativeness and quality of the data pieces to be purchased. Based on these estimated information, \system recommends the data pieces on the data marketplace that obtain the best correlation while satisfy the given constraints. 
In particular, We make the following contributions. 

\begin{figure*}[h!]
\begin{center}
\includegraphics[width=0.3\textwidth]{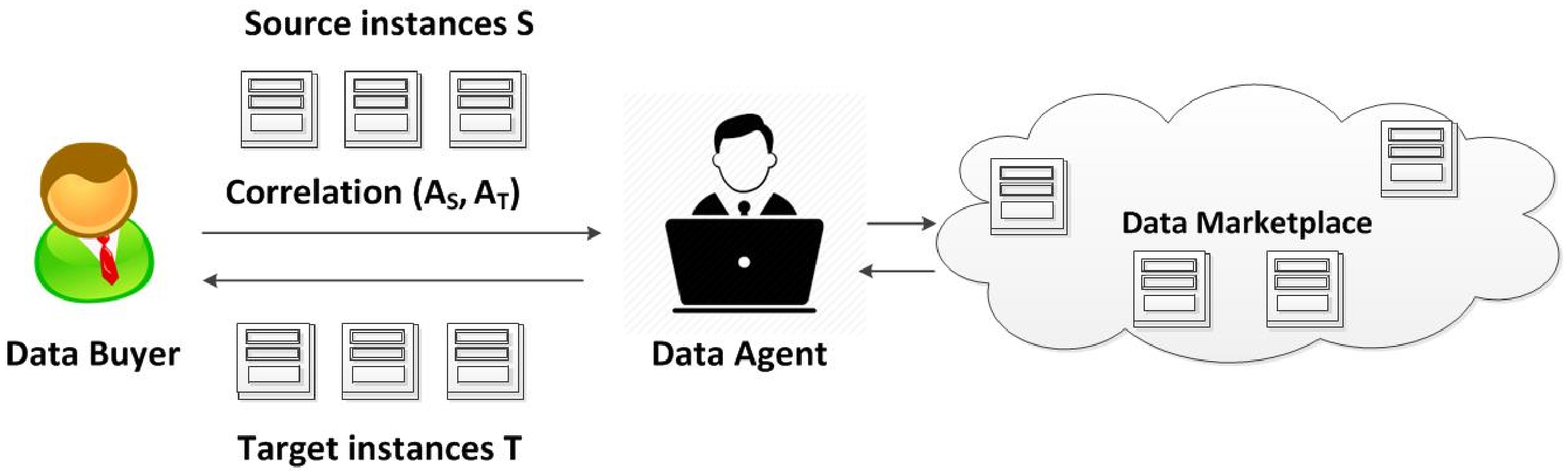}
\caption{\label{fig:model} Framework of \system}
\end{center}
\vspace{-0.1in}
\end{figure*}

First, we design a new sampling method named {\em correlated re-sampling} to deal with multi-table joins of large intermediate join size. We show how to unbiasedly estimate the correlation, join informativeness, and quality based on the samples. 

Second, we design a new two-layer {\em join graph} structure. The join graph is constructed from the samples collected from the data marketplace. It includes the join relationship of the data instances at schema level, as well as the information of correlation, join informativeness, quality, and price of the data at instance level.

Based on the join graph, third, we transform the data acquisition problem to a graph search problem, which searches for the optimal subgraph in the join graph that maximizes the (estimated) correlation between source and target nodes, while the quality, join informativeness (in format of edge weights in the graph), and price satisfy the specified constraints. We prove that the problem of searching for such optimal subgraph is NP-hard. Thus, we design a heuristic algorithm based on
Markov chain Monte Carlo (MCMC). Our heuristic algorithm first searches for the minimal weighted graph at the instance layer of the join graph, which corresponds to a set of specific instances. Then it finds the optimal target graph at the attribute set layer of the join graph, which corresponds to the attributes in those instances that are to be purchased from the marketplace. 

Last but not least, we perform an extensive set of experiments on large-scale benchmark datasets. The result show that our search method can find the acquisition results with high correlation efficiently. 

The rest of the paper is organized as following. Section \ref{sc:pre} introduces the preliminaries. 
Section \ref{sc:estimation} presents the data sampling method.
Section \ref{sc:jgraph} and 
Section \ref{sc:algo}  present the off-line and online phase of \system respectively.
 Section \ref{sc:exp} shows the experiment results. 
Section \ref{sc:related} discusses the related work.
Section \ref{sc:conclusion} concludes the paper.

\vspace{-0.05in}
\section{Preliminaries}
\label{sc:pre}
 
\nop{
and introduce the notation used throughout the paper.
We use notations in Table \ref{table:notation}.

\begin{table}
\label{table:notation}
\begin{center}
\begin{tabular}{|c|c|} \hline
notation & meaning \\\hline
$G$ & Join graph \\\hline
$V$ & Set of nodes of graph $G$ \\\hline
$E$ & Set of edges of graph $G$ \\\hline
$D$ & Original instance\\\hline
$R$ & Purchased instance (a subset of $D$)\\\hline
$A$ & An attribute \\\hline
$D_S$ & Source instance\\\hline
$\Set{T}$ & Target attributes\\\hline
$\mathcal{L}$ & Attribute set lattice \\\hline
$\mathcal{J}$  & Join result \\\hline
$T$  & A tree from graph $G$ \\\hline
$J$  & Join attribute \\\hline
$p()$ & Pricing function \\\hline
$Q$ & Date quality \\\hline
$JS$ & Join strength \\\hline
$F()$ & Objective function \\\hline
\end{tabular}
\caption{Notations}
\end{center}
\end{table}
}

\subsection{System Overview}

We consider a data marketplace $\mathbf{M}$ that stores and maintains a collection of relational database instances $\mathcal{D}$ = $\{D_1, \\ \dots, D_n\}$. 
A {\em data shopper} may own a set of relational data instances $\Set{S}$ locally, which contain a set of {\em source attributes} $\mathcal{A_S}$. 
The data shopper would like to purchase a set of {\em target attributes} $\mathcal{A_T}$ from $\mathbf{M}$, so that the correlation between $\mathcal{A_S}$ and $\mathcal{A_T}$ is maximized. In this paper, we only consider {\em vertical purchase}, i.e., the shopper buys the attributes from the data marketplaces.
We assume the data shopper fully trusts \system. Figure \ref{fig:model} illustrates the framework of \system.  \system consists of two phases: the {\em offline} and {\em online} phases.

{\bf Offline phase.} During  the offline phase, \system collects a set of samples from $\mathbf{M}$, and constructs a data structure named {\em join graph} that includes: (1)  the join relationship among the samples at schema level; and (2) the information of correlation, join informativeness, quality, and price at instance level. Due to the high complexity of calculating the exact correlation, data quality, and join informativeness from the join results, the correlation, quality, and join informativeness are estimated from the collected samples.

{\bf Online phase.} During the online phase, \system accepts the acquisition requests from data shoppers. Each acquisition request consists of: (1) the source dataset $\Set{S}$ that is owned by the shopper and the source attributes $\mathcal{A_S}\in\Set{S}$; (2) 
the target attributes $\mathcal{A_T}\in \mathcal{D}$,  and (3) the budget $B$ of data purchase from $\mathbf{M}$, to \system. The data shopper can also specify his requirement of data quality and join informativeness by the setup of the corresponding threshold values. 
Both $\Set{S}$ and $\mathcal{A_S}$ are optional. The acquisition without $\Set{S}$ and $\mathcal{A_S}$ aims to find the best correlation of $\mathcal{A_T}$ in $\mathcal{D}$. 

After \system receives the acquisition request, it first processes the request on its local join graph by searching for a set of {\em target instances} $\Set{T} =\{T_1, \dots, T_k\}$, where: 
(1) for each $T_i$, there exists an instance $D_j \in \mathcal{D}$ such that $T_i\subseteq D_j$; 
(2) the correlation between $\mathcal{A_S}$ and $\mathcal{A_T}$ is maximized in $\mathcal{J} = \bowtie_{\forall T_i\in \Set{S} \cup \Set{T}}T_i$, where $\bowtie$ denotes the equi-join operator; 
(3) each $T_i\in\Set{T}$ is assigned a {\em price}, which is decided by a query-based pricing function \cite{balazinska2011data}. The total price of $\Set{T}$ should not exceed $B$; 
and (4) the quality and join informativeness of $\mathcal{J}$  are beyond  the threshold specified by the data shopper. 
If no such target instance can be identified from its current join graph and data samples,
\system purchases more samples from $\mathbf{M}$, updates its local join graph, and performs the data search again. The iterative process continues until either the desired $\Set{T}$ is found or the data shopper changes his acquisition requirement (e.g., by relaxing the data quality threshold). The data shopper pays \system for the sample purchase during acquisition.

After \system identifies the target instances $\Set{T}$, it generates a set of SQL projection queries $\mathcal{Q}$, where each instance $T_i\in\Set{T}$ corresponds to a query $\mathsf{Q}\in\mathcal{Q}$.  Each $\mathsf{Q}\in\mathcal{Q}$, denoted as $\pi_{A_i}(D_i)$ $(\pi:$ the projection operator), selects a set of attributes $A_i$ from the instance $D_i$. We call the attribute set $A_i$ the {\em projection} attribute set.  In the paper, we use $D.A$ to specify the values of attribute $A$ of the instance $D$. 

After receiving the purchase option $\mathcal{Q}$ from \system, the data shopper sends $\mathcal{Q}$ to $\mathbf{M}$ directly, and obtains the corresponding instances. 

\subsection{Data Quality}
\label{sc:dquality}

In this paper, we consider the setting where data on the marketplaces is {\em dirty}. In this project, we mainly consider one specific type of dirty data, the {\em inconsistent} data, i.e., the data items that violate  integrity constraints \cite{arenas1999consistent,rahm2000data}. 
A large number of existing works (e.g., \cite{fan2008conditional,chiang2008discovering}) specify the data consistency in the format of {\em functional dependency ($FD$)}. 
Formally, given a relational dataset $D$, a set of attributes $Y$ of $D$ is said to be {\em functionally dependent} on another set of attributes $X$ of $D$ (denoted $X\rightarrow Y$) if and only if for any pair of records $r_1, r_2$ in $D$, if $r_1[X] =r_2[X]$, then $r_1[Y]=r_2[Y]$. 
For any FD $F: X \rightarrow Y$ where $Y$ contains multiple attributes, $F$ can be decomposed in to multiple FD rules $F': X\rightarrow Y'$, with $Y'$ containing a single attribute of $Y$. Thus for the following discussions, we only consider FDs that contain a single attribute at the right hand side. 

Based on FDs, the data inconsistency can be measured as the amounts of data that does not satisfy the FDs. Before we formally define the measurement of data inconsistency, first, we formally define {\em partitions}. We use $r.A$ to denote the value of attribute $A$ of record $r$. 
\begin{nameddefinition}{Equivalence Class ($EC$) and Partitions}\cite{huhtala1999tane}
\label{df:ec}
The {\em equivalence class} ($EC$) of a tuple $r$ with respect to an attribute set $X$, denoted as $eq_X^r$, is defined as $eq_X^r = \{r'| r.A =r'.A, \forall A\in X, \forall r'\in D\}$. 
The set $\pi_X = \{eq_X^r|r\in D\}$ is defined as a {\em partition} of $D$ of the attribute set $X$. Informally, $\pi_X$ is a collection of disjoint sets ($ECs$) of tuples, such that each set has a unique representative value of the set of attributes $X$, and the union of the sets equals $D$. 
\end{nameddefinition}

Given a dataset $D$ and an FD $F: X \rightarrow Y$, where $X$ and $Y$ are a set of attributes, the degree of data inconsistency by $F$ can be measured by the fraction of records that do not satisfy FDs. Following this, 
we define the data quality function $Q(D, F)$ for a given FD $F$ on an instance $D$. 

\begin{nameddefinition}{Data quality of one instance w.r.t. one FD}
\label{def:q1s1fd}
Given a data instance $D$ and a FD $F: X\rightarrow Y$ on $D$, for any equivalence class $eq_x\in \pi_{X}$, the {\em correct equivalence class} in $\pi_{X\cup Y}$ is 
\vspace{-0.05in}
\begin{equation}
\begin{aligned}
\label{eqn:e1c1}
C(D, X\rightarrow Y, eq_x) = & \{eq_{xy}|eq_{xy}\in \pi_{X\cup Y}, eq_{xy}\subseteq eq_x,  \\ 
& \nexists eq'_{xy}\in  \pi_{X\cup Y} \text{ } s.t. \text{ } eq'_{xy}\subseteq eq_x \text{ and } \\
& |eq_{xy}'|>|eq_{xy}|\}.
\end{aligned}
\end{equation}
In other words, $C(D, X\rightarrow Y, eq_x)$ is the largest equivalence class in $\pi_{XY}$ with the same value on $X$ attributes as $eq_x$. 
If there are multiple such equivalence classes of the same size, we break the tie randomly.

Based on the definition of correct equivalence classes, the set of correct records in $D$ w.r.t. $F$ is defined as:
\vspace{-0.05in}
\begin{equation}
\label{eqn:c1f1}
C(D, X\rightarrow Y) = \bigcup_{eq_x\in \pi_{X}} C(D, X\rightarrow Y, eq_x).
\end{equation}


The quality of a given data instance $D$ w.r.t. $F:X\rightarrow Y$ is measured as:
\begin{equation}
\label{eqn:q1f1c}
Q(D, X \rightarrow Y) = \frac{|C(D, X\rightarrow Y)|}{|D|}, 
\end{equation}
\end{nameddefinition}

\begin{table}[!htbp]
\begin{center}
\begin{small}
\begin{tabular}{|c|c|c|}
\hline
TID & A & B \\\hline
$t_1$ & $a_1$ & $b_1$ \\\hline
$t_2$ & $a_1$ & $b_1$ \\\hline
$t_3$ & $a_1$ & $b_2$ \\\hline
$t_4$ & $a_1$ & $b_3$ \\\hline
$t_5$ & $a_2$ & $b_2$ \\\hline
\end{tabular}
\end{small}
\caption{\label{tb:1f1d}An example of data instance $D$ ($FD: A\rightarrow B$)}
\end{center}
\end{table}

\begin{example}
Consider the data instance $D$ in Table \ref{tb:1f1d} and the FD $A\rightarrow B$ on $D$. The partition $\pi_{A}$ on $A$ consists of two equivalence classes, $eq_{A}^{a_1}=\{t_1, t_2, t_3, t_4\}$ and $eq_A^{a_2}=\{t_5\}$. The partition $\pi_{AB}$ consists of four equivalence classes, i.e., $eq_{AB}^{a_1,b_1}=\{t_1, t_2\}$, $eq_{AB}^{a_1,b_2}=\{t_3\}$, $eq_{AB}^{a_1,b_3}=\{t_4\}$, and $eq_{AB}^{a_2,b_2}=\{t_5\}$. Then $C(D, A\rightarrow B, eq_{A}^{a_1})=eq_{AB}^{a_1,b_1}$, while $C(D, A\rightarrow B, eq_{A}^{a_2})=eq_{AB}^{a_2,b_2}$. Therefore, the set of correct tuples is $C(D, A\rightarrow B)=\{t_1, t_2, t_5\}$. The tuples $t_3$ and $t_4$ are considered as error. 
\end{example} 

\nop{
\begin{nameddefinition}{Data quality of one instance}
\label{def:q1smfd}
Given an instance $D$ and a set of AFDs $\mathcal{F}$ on $D$, the set of correct records of $D$ w.r.t. $\mathcal{F}$ are defined as:
\nop{\begin{equation}
\label{eqn:e1dmf}
E(D, \mathcal{F}) = \bigcup_{\forall F_i\in\mathcal{F}} E(D, F_i),
\end{equation}}
\begin{equation}
\label{eqn:c1dmf}
C(D, \mathcal{F}) = \bigcap_{\forall F_i\in\mathcal{F}} C(D, F_i),
\end{equation}
where $C(D, F_i)$ follows Equation \ref{eqn:c1f1}.
 And the quality of $D$ w.r.t. $\mathcal{F}$ is measured as
 \begin{equation}
 \label{eqn:q1dmf}
Q(D)=\frac{|U|}{|D|}, 
\end{equation}
where $U = C(D, \mathcal{F})$ (Equation \ref{eqn:c1dmf}).
\end{nameddefinition}
}
As we consider the dirty datasets that may be inconsistent, the exact FDs may not hold on those dirty datasets. Instead we consider {\em approximate functional dependencies} (AFDs) which allow FDs to hold on the data with small errors. Formally, an AFD $F$ holds on the dataset $D$ if $Q(D, F)\geq \theta$, where $\theta\in (0,1)$ is a user-defined threshold value. 
Based on the definition of AFDs, now we are ready to define the quality of a set of instances. 

\begin{nameddefinition}{Data quality}
\label{def:qmsmfd}
Given a set of instances $\mathcal{D}$, let $\mathcal{J}$ = $\bowtie_{\forall D_i\in\mathcal{D}} D_i$, and $\mathcal{F}$ be the set of AFDs hold on $\mathcal{J}$. 
The set of correct records of $D$ w.r.t. $\mathcal{F}$ are defined as
$C(\mathcal{J}, \mathcal{F}) = \bigcap_{\forall F_i\in\mathcal{F}} C(\mathcal{J}, F_i)$, where $C(\mathcal{J}, F_i)$ follows Equation (\ref{eqn:c1f1}).
\nop{
\begin{equation}
\label{eqn:c1dmf}
C(\mathcal{J}, \mathcal{F}) = \bigcap_{\forall F_i\in\mathcal{F}} C(\mathcal{J}, F_i)
\end{equation}
where $C(\mathcal{J}, F_i)$ follows Equation (\ref{eqn:c1f1}).
}
The quality of $\mathcal{D}$ w.r.t. $\mathcal{F}$ is measured as
\begin{equation}
\label{eqn:qmdmf}
Q(\mathcal{D})=\frac{|C(\mathcal{J}, \mathcal{F})|}{|\mathcal{J}|},
\end{equation}
\end{nameddefinition}
Intuitively, the quality is measured as the portion of the records that are correct among all the FDs on the instance. 


\noindent{\bf Impact of Join on Data Quality.} 
To provide high-quality data for purchase, a naive solution is to clean the inconsistent data off-line before processing any data acquisition request online. However, this solution in incorrect, since {\em join can change data quality}. High-quality datasets may become low-quality after join, and vice versa. Example \ref{exp:join} shows an example. 

\begin{table}[!htbp]
\begin{small}
\begin{center}
\begin{tabular}{|c|c|c|c|}
\hline
TID & A & B & C \\\hline
$t_1$ & $a_1$ & $b_1$ & $c_4$ \\\hline
$t_2$ & $a_1$ & $b_1$ & $c_5$ \\\hline
\vdots & \vdots & \vdots & \vdots \\\hline
$t_{996}$ & $a_1$ & $b_1$ & $c_{999}$ \\\hline
$t_{997}$ & $a_1$ & $b_2$ & $c_{1}$ \\\hline
$t_{998}$ & $a_1$ & $b_2$ & $c_2$ \\\hline
$t_{999}$ & $a_1$ & $b_3$ & $c_3$ \\\hline
$t_{1000}$ & $a_1$ & $b_3$ & $c_3$ \\\hline
\end{tabular}
\\
(a) $D_1: A \rightarrow B$ ($Q(D_1)=0.996$)
\\
\begin{tabular}{cc}
\begin{tabular}{|c|c|c|c|}
\hline
TID & C & D & E \\\hline
$t_1$ & $c_1$ & $d_1$ & $e_1$ \\\hline
$t_2$ & $c_1$ & $d_1$ & $e_1$ \\\hline
$t_3$ & $c_2$ & $d_1$ & $e_2$ \\\hline
$t_4$ & $c_3$ & $d_1$ & $e_2$ \\\hline
$t_5$ & $c_4$ & $d_1$ & $e_2$ \\\hline
\end{tabular}
&
\begin{tabular}{|c|c|c|c|c|c|}
\hline
TID & A & B & C & D & E \\\hline
$t_1$ & $a_1$ & $b_2$ & $c_1$ & $d_1$ & $e_1$ \\\hline
$t_2$ & $a_1$ & $b_2$ & $c_1$ & $d_1$ & $e_1$ \\\hline
$t_3$ & $a_1$ & $b_2$ & $c_2$ & $d_1$ & $e_2$ \\\hline
$t_4$ & $a_1$ & $b_3$ & $c_3$ & $d_1$ & $e_2$ \\\hline
$t_5$ & $a_1$ & $b_3$ & $c_3$ & $d_1$ & $e_2$ \\\hline
\end{tabular}
\\
(b) $D_2: D \rightarrow E$
& 
(c) $D_1\Join D_2$: $A\rightarrow B$, $D\rightarrow E$
\\
$Q(D_2)=0.6$
&
$Q(D_1\Join D_2)=0.2$
\end{tabular}
\end{center}
\end{small}
\caption{\label{tb:join} An example that the join of two high-quality data instances is low-quality}
\vspace{-0.05in}
\end{table}

\begin{example}
\label{exp:join}
In Table \ref{tb:join} (a) and (b), we display two datasets with good quality. The correct records in $D_1$ include $\{t_1, t_2, \dots, t_{996}\}$. Thus $Q(D_1)=0.996$. Similarly, the correct records in $D_2$ include $\{t_3, t_4, t_5\}$. Thus $Q(D_2)=0.6$. 
After the join of $D_1$ and $D_2$  (the join result is shown in Table \ref{tb:join} (c)), there only exist 5 tuples. We get $C(D_1\Join D_2, A\rightarrow B)=eq_{(a_1, b_2)}=\{t_1, t_2, t_3\}$, while $C(D_1\Join D_2, D\rightarrow E)=eq_{(d_1, e_2)}=\{t_3, t_4, t_5\}$. Therefore, $Q(D_1\Join D_2)=0.2$. 
\end{example}

Example \ref{exp:join} shows that performing data cleaning before join is not acceptable. The data quality has to be measured on the join result. 

\nop{
\subsection{Join Strength}
Given two data instances $D_1$ and $D_2$, joining $D_1$ with different subsets of $D_2$ delivers different results. 
For instance, consider $D_1$ of schema $(A, B, C)$ and $D_2$ of schema $(A, B, D)$. Alice can purchase $\pi_{A}(D_2)$, or $\pi_{B}(D_2)$, or $\pi_{A, B}(D_2)$ to join with $D_1$. Picking which purchase option is decided by which returns more join results. Next, we formally define the {\em join strength} of a set of instances. 

\begin{nameddefinition}{Join Strength}
\label{def:js}
Given a set of instances $\mathcal{D}$ = $\{D_1, \dots, D_k\}$, let $ \mathcal{J}$ = $\bowtie_{i=1}^{k} D_i$. The join strength of $\mathcal{D}$ is measured as:
\begin{equation}
JS(\mathcal{D})=\frac{|\mathcal{J}|}{\prod_{i=1}^{k}|D_i|}.
\end{equation}
\end{nameddefinition}
}

\subsection{Join Informativeness}
Evaluating whether the join result is informative simply by join size is not suitable. For example, consider the join of $D_S$ and $D_5$ in Table 1. The join is meaningless since it associates the aggregation data with individual records. But its join size is larger than the other options (e.g., join $D_S$ with $D_1$) that are more meaningful. Therefore, we follow the information-theory
measures in \cite{yang2011summary} to quantify how informative the join is. Given two attribute sets $X$ and $Y$, 
let $I(X, Y)$ and $H(X, Y)$ denote the mutual information and 
entropy of the joint distribution $(X, Y)$ respectively. The join informativeness is formally defined below. 
\begin{nameddefinition}{Join Informativeness \cite{yang2011summary}}
Given two instances $D$ and $D'$, let $J$ be their join attribute(s). 
The join informativeness of $D$ and $D'$ is defined as 
\begin{equation}
\label{eqn:ji}
JI(D, D') = \frac{H(D.J, D'.J) - I(D.J, D'.J)}{H(D.J, D'.J)},
\end{equation}
by using the joint distribution of $D.J$ and $D'.J$ in the output of the full outer join of $D$ and $D'$. 
\end{nameddefinition}
The reason of using the outer join for the join informativeness measurement is to deal with values in $D$ and $D'$ that
do not match, i.e., pairs of type (val, NULL). The informativeness 
function penalizes those joins with excessive
numbers of such unmatched values \cite{yang2009summarizing}. 
The join informativeness value always falls in the range [0, 1]. 
It is worth noting that the smaller $JI(D, D', J)$ is, the more important is the join connection between $D$ and $D'$.  

\subsection{Correlation Measurement}
\label{sc:corr}
Quite a few correlation measurement functions (e.g. {\em Pearson correlation coefficient} , {\em token inverted correlation} \cite{song2010efficient}, and {\em cross correlation (autocorrelation)}) can evaluate the correlation of multiple attributes. However, they can deal with either categorical or numerical data, but not both. Therefore, we use the Shannon entropy based correlation measure \cite{nguyen2014detecting} to quantify the correlation as it deals with both categorical and numerical data. 

\begin{nameddefinition}{Correlation \cite{nguyen2014detecting}}
Given a dataset $D$ and two attribute sets $X$ and $Y$, the {\em correlation} of $X$ and $Y$ $CORR(X, Y)$ is measured as 
\begin{itemize}
\item $CORR(X, Y) = H(X) - H(X|Y)$ if $X$ is categorical,
\item $CORR(X, Y) = h(X) - h(X|Y)$ if $X$ is numerical,
\end{itemize}
where $H(X)$ is the Shannon entropy of $X$ in $D$, and $H(X|Y)$ is the conditional entropy:  
\[H(X|Y ) = \sum H(X|y)p(y),\]
Furthermore, $h(X)$ is the cumulative entropy of attribute $X$
\[h(X) = -\int P(X\leq x)log P(X\leq x)dx,\]
and 
\[h(X|Y) = -\int h(X|y)p(y)dy.\]
\end{nameddefinition}

\subsection{Problem Definition}
  \nop{This is a multi-objective optimization problem. We scalarize the two objectives (maximize quality and join strength) into a single objective by adding each of the objective function with user-specified weights. We assume that the weight $\alpha$ is associated with the quality, and thus (1 - $\alpha$) with the join strength. Accordingly, we define the single objective function that combines both data quality and join strength as:
Accordingly, we define the objective function as:
\begin{equation}
\label{def:objfunction}
F(\mathcal{R}) = \alpha Q(\mathcal{R}) + (1 - \alpha)JS(\mathcal{R}), 
\end{equation}
where $\mathcal{R}$ is a set of given instances. 
}

Intuitively, the data shopper prefers to purchase the high-quality, affordable data instances that can deliver good join informativeness, and most importantly, the best correlation. 
We assume the data samples have been obtained from the data marketplace. 
Formally, given a set of data samples $\mathcal{D}=\{D_1, \dots, D_n\}$ collected from the data marketplace, a set of source instances $\Set{S}$ with source attributes $\mathcal{A_S}$, a set of target attributes $\mathcal{A_T}$, and the budget $B$ for data purchase, the data acquisition problem is defined as the following: 
find a set of database instances $\Set{T} \subseteq \mathcal{D}$ such that
\begin{equation*}
\begin{aligned}
& \underset{\Set{T}}{\text{maximize}}
& & CORR(\mathcal{A_S}, \mathcal{A_T})\\
& \text{subject to}
& &\forall T_i\in\Set{T}, \exists D_j \in \mathcal{D}\ s.t.\ T_i\subseteq D_j,\\
&&& \sum_{T_i\in\Set{S}\cup\Set{T}}JI(T_i, T_{i+1}) \leq \alpha, \\
&&& Q(\Set{T}) \geq \beta,\\
&&&  p(\Set{T})\leq B, 
\end{aligned}
\end{equation*}
where $\alpha$ and $\beta$ are the user-specified thresholds for join informativeness and quality respectively. 
We note that $CORR(\mathcal{A_S}, \mathcal{A_T})$ is measured on the join result. 
We assume that the data shopper can be guided to specify appropriate parameter values for $\alpha$ and $\beta$.
The output $\Set{T}$ will guide \system to suggest the data shopper of the SQL queries for data purchase from the data marketplaces. 

\section{Data Sampling and Estimation}
\label{sc:estimation}
In this section, we discuss how to estimate join informativeness,  data quality, and correlation based on samples. 
We use the correlated sampling \cite{vengerov2015join} method to  generate the samples from the data marketplace. General speaking, for any tuple $t_i\in D_1$, let $t_i[J]$ be its join attribute value. The record $t_i$ is included in the sample $S_1$ if $h(t_i[J])\leq p_1$, where $h()$ is the hash function that maps the join attribute values uniformly into the range $[0,1]$, and $p_1$ is the sampling rate. 
The sample $S_2$ of $D_2$ is generated in the same fashion. 

\subsection{Estimation of Join Informativeness} 
For a pair of data instances $D_1$ and $D_2$, let $S_1$ and $S_2$ be the samples of $D_1$ and $D_2$ by using correlated sampling. Then the join informativeness $JI(D_1,D_2)$ is estimated as:
\begin{equation}
\label{eqn:eji}
\hat{JI}(D_1,D_2)=JI(S_1, S_2),
\end{equation}
where $JI()$ follows Equation (\ref{eqn:ji}).

Next, we have Theorem \ref{th:ji} to show that the estimated join informativeness is unbiased and is expected to be accurate. 
\begin{theorem}
\label{th:ji}
Let $D_1$ and $D_2$ be two data instances, and $S_1$ and $S_2$ be the sampled dataset by using correlated sampling with the same sampling rate. It must be true that the expected value of the estimated join informativeness $E(JI(S_1,S_2))$ satisfies that:
\[E(JI(S_1,S_2))=JI(D_1, D_2).\]
\end{theorem}
We present the proof of Theorem \ref{th:ji} in our full paper \cite{cost2018li}.

\subsection{Correlated Re-sampling for Estimation of Correlation \& Quality}

One weakness of the correlated sampling for estimation of the correlation and data quality is that the size of join result from samples can be extremely large, especially for the join of a large number of data instances. Note that this is not a problem for estimation of join informativeness as it only deals with 2-table joins.  
To deal with the large join result, we design the {\em correlated re-sampling} method by adding a second-round sampling of the join results. Intuitively, given a set of data instances $(D_1, \dots, D_p)$, for any intermediate join result $IJ$, if its size exceeds a user-specified threshold $\eta$, we sample$IJ'$ from $IJ$ by using a fixed re-sampling rate, and use $IJ'$ for the following joins. In this way, the size of the intermediate join result is bounded.

For the sake of simplicity, for now we only focus on the re-sampling of 3-table joins (i.e., $D_1\Join D_2 \Join D_3$). Let $S_1$, $S_2$ and $S_3$ be the samples of $D_1$, $D_2$, and $D_3$ respectively. Let $S_{12}'$ denote the re-sampling result of $S_1\Join S_2$ if its size exceeds $\eta$, or $S_1\Join S_2$ otherwise. The estimated correlation and quality are 
\begin{equation}
\label{eq:ec}
\widehat{CORR}_{D_1\Join D_2 \Join D_3}(\mathcal{A_S},\mathcal{A_T})=CORR_{S_{1,2}'\Join S_3}(\mathcal{A_S},\mathcal{A_T}),
\end{equation}
\begin{equation}
\label{eq:eq}
\hat{Q}(D_1, D_2, D_3)=Q(S_{1,2}', S_3).
\end{equation}

The estimation can be easily extended to the join paths of arbitrary length, by applying  sampling on the intermediate join results. Next, we show that the correlation and data quality estimation by using correlated re-sampling is also unbiased.

\begin{theorem}
\label{th:cq}
Given a join path $(D_1, D_2, D_3)$, let $S_{1,2}'$ be the sample from $S_1 \Join S_2$. It must be true that the expected value of the estimated correlation $E(CORR_{S_{1,2}'\Join S_3})$ satisfies:
\[E(CORR_{S_{1,2}'\Join S_3}(\mathcal{A_S},\mathcal{A_T}))=CORR_{D_1\Join D_2 \Join D_3}(\mathcal{A_S},\mathcal{A_T}),\] 
and the expected value of the estimated quality $E(Q({S_{1,2}', S_3}))$ satisfies:
\[E(Q({S_{1,2}', S_3}))=Q(D_1, D_2, D_3),\] 
where $\mathcal{A_S}$ and $\mathcal{A_T}$ are the source and target attribute sets that are included in $D_1$ and $D_3$.
\end{theorem}

Due to the space limit, we show the proof of Theorem \ref{th:cq} in our full paper \cite{cost2018li}. 
We must note that the estimation is unbiased regardless of the value of $\eta$.

\nop{
\noindent{\bf Correlation estimation.} 
Given two data instances $T_1$ and $T_2$, let $J$ be their join attribute(s). For the sake of simplicity, assume that $J$ only includes one attribute. Let $s_1$ and $s_2$ be the desired size of samples from $T_1$ and $T_2$ respectively. Correlated sampling \cite{vengerov2015join} produces the samples in the following way.
For any tuple $t_i\in T_1$, let $t_i[J]$ be its join attribute value. Then $t_i$ is included in the sample $S_1$ if $h(t_i[J])\leq p_1$, where $h()$ is the hash function that maps the join attribute values uniformly into the range $[0,1]$, and $p_1=\frac{s_1}{|T_1|}$. 
The sample $S_2$ of $T_2$ is generated in the same fashion.
We have Theorem \ref{th:correlated} to show that the estimated correlation from the samples is unbiased. 

\begin{theorem}
\label{th:correlated}
Let $T_1$ and $T_2$ be two data instances, $S_1$ and $S_2$ be the sampled dataset from $T_1$ and $T_2$ following correlated sampling. It must be true that 
\[E(CORR_{S_1\Join S_2} (X,Y))=CORR_{T_1\Join T_2}(X,Y),\] 
where $X$ and $Y$ are the source and target attribute in $T_1$ and $T_2$. 
\end{theorem}

We can easily extend Theorem \ref{th:correlated} to prove that the estimated correlation of multi-join from samples (i.e., joining more than two data instances) is still unbiased. We omit the details due to the space limit.

\Wendy{move proof to Appendix.}

\noindent{\bf Correlated resampling.} 

\Wendy{ }
}

\section{Off-line Phase: Construction of Join Graph}
\label{sc:jgraph}

In this section, we present the concept of {\em join graph}, the main data structure that will be used for our search algorithm (Section \ref{sc:algo}). 
First, we define the {\em attribute set lattice} of a single data instance. 
\begin{nameddefinition}{Attribute Set Lattice (AS-lattice)}
\label{def:lattice}
Given a data instance $D$ that has $m$ attributes $\mathcal{A}$, it corresponds to an {\em attribute set lattice} (AS-lattice) $\mathcal{L}$, in which each vertex corresponds to a unique attribute set $\mathcal{A}'\subseteq \mathcal{A}$. For the vertex $v\in\mathcal{L}$ whose corresponding attribute set is $\mathcal{A}'$, it corresponds to the projection instance  $\pi_\mathcal{A'}(D)$.  
Given two lattice vertice $L_1, L_2$ of $\mathcal{L}$, $L_2$ is $L_1$'s {\em child} (and linked by an edge) if (1) $\mathcal{A}_1\subseteq \mathcal{A}_2$, and (2) $|\mathcal{A}_2|=|\mathcal{A}_1|+1$, where $\mathcal{A}_1$ and $\mathcal{A}_2$ are the corresponding attribute sets of $L_1$ and $L_2$. $L_1$ is an {\em ancestor} of $L_2$ (linked by a path in $\mathcal{L}$) if $\mathcal{A}_1\subset \mathcal{A}_2$. 
$L_1$ and $L_2$ are {\em siblings} if they are at the same level of $\mathcal{L}$.  
\end{nameddefinition}

\begin{figure}
\begin{center}
\includegraphics[width=0.4\textwidth]{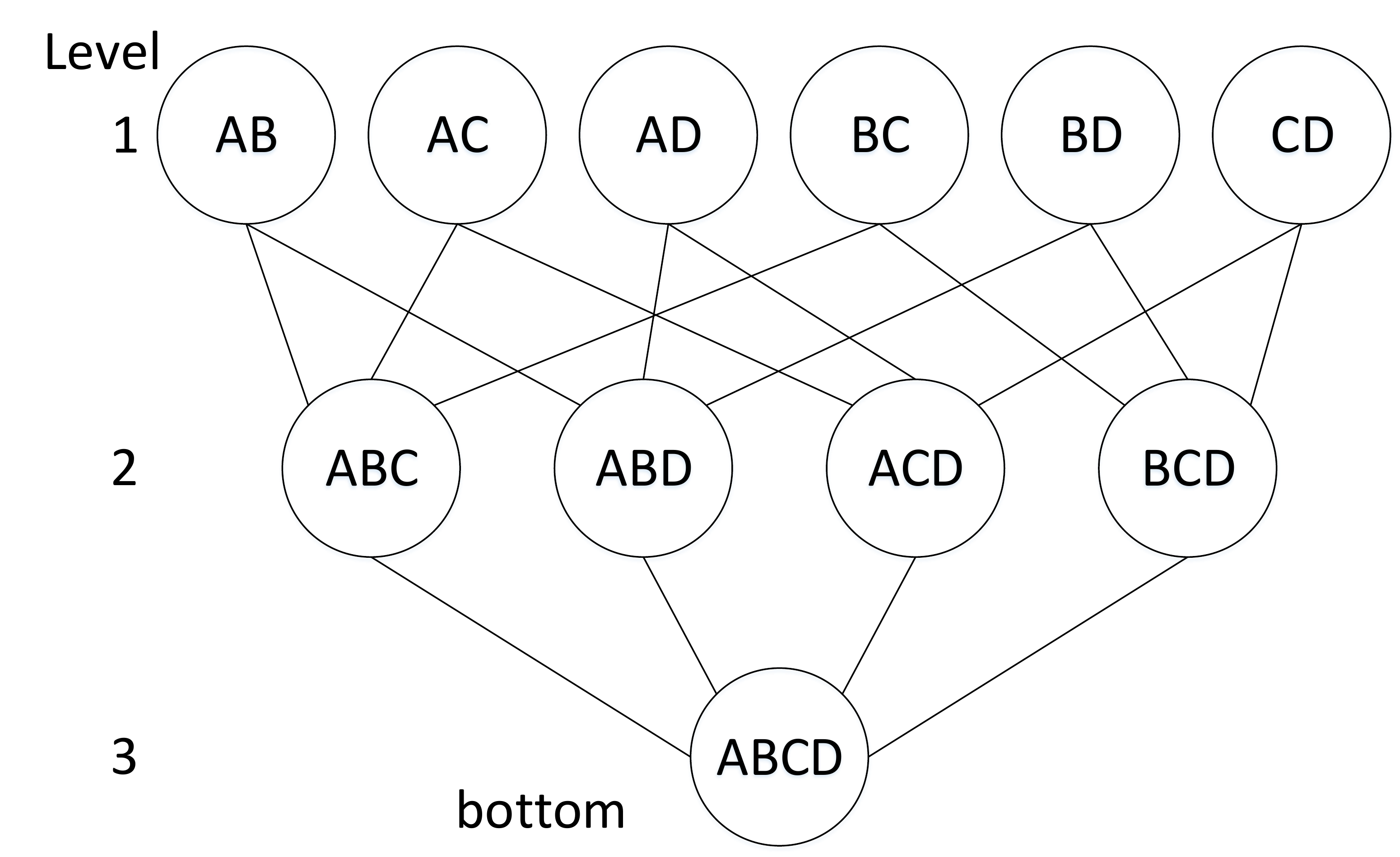}
\caption{\label{fig:lattice} An example of attribute set lattice}
\vspace{-0.15in}
\end{center}
\end{figure}

Given $D$ and its $m$ attributes $\mathcal{A}$, the height of the attribute set lattice of $D$ is $m-1$. The bottom of the lattice contains a single vertex, which corresponds to $\mathcal{A}$, while the top of the lattice contains ${m}\choose{2}$ vertices, each corresponding to a unique 2-attribute set. Figure \ref{fig:lattice} shows an example of the attribute set lattice of an instance of four attributes $\{A, B, C, D\}$. In general, given an instance of $m$ attributes, its attribute set lattice consists of ${m}\choose{2}$ + $\dots$ + ${m}\choose{m}$ = $2^m - m - 1$ vertices. 

\begin{figure*}
\begin{center}
\includegraphics[width=0.65\textwidth]{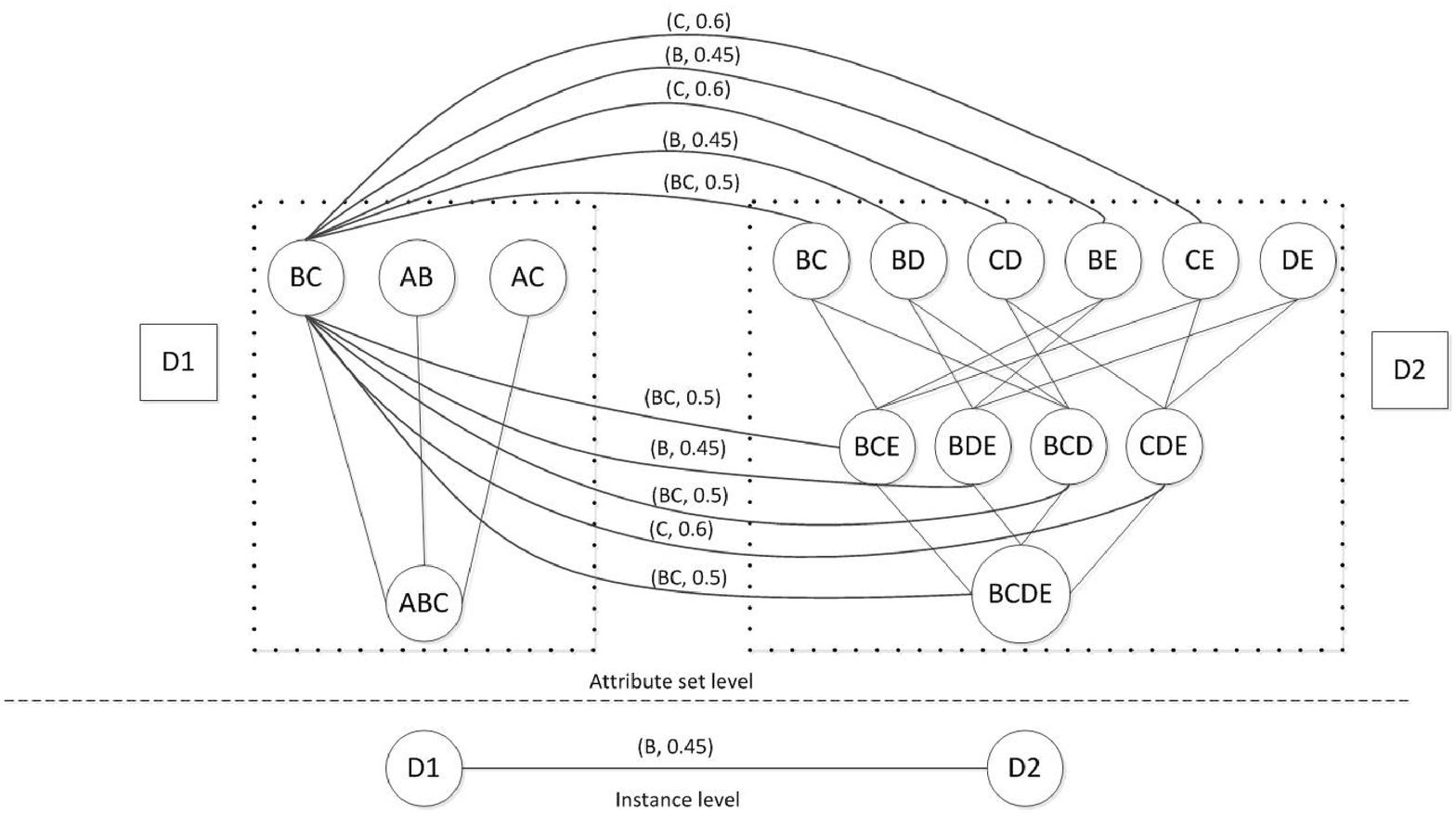}
\caption{\label{fig:jgraph}An example of a join graph for two instances $D_1(ABC)$ and $D_2(BCDE)$. Dotted rectangles represent 
tables. Only the node $BC$ of instance $D_1$ has all edges. The edges of other nodes are omitted for simplicity. }
\end{center}
\vspace{-0.1in}
\end{figure*}

Next, we define {\em join graph}. In the following discussion, we use $\mathsf{AS}(v)$ to denote the attribute set that the vertex $v$ corresponds to. 

\begin{nameddefinition}{Join Graph}
\label{def:jgraph}
Given a set of data instances (samples) $\mathcal{D} = \{D_1, \dots, D_n\}$, it corresponds to an undirected, weighted, two-layer {\em join graph} $G$:
\begin{itemize}
\item {\bf Instance layer (I-layer)}: each vertex at this layer, called {\em instance vertex} (I-vertex), represents a data instance $D_i \in\mathcal{D}$. There is an edge $e_{ij}$, called {\em I-edge}, between any two I-vertices $v_i$ and $v_j$ if $\mathsf{AS}(v_i)\cap \mathsf{AS}(v_j)\neq\emptyset$. 
\item {\bf Attribute set layer (AS-layer)}: for each I-vertex $v_i$, it projects to a set of vertices, called {\em attribute set} vertex (AS-vertex), at the AS-layer, each corresponding to an unique attribute set of the  instance $D_i$. 
The AS-vertices that was projected by the same I-vertex construct the AS-lattice (Def. \ref{def:lattice}). Each AS-vertex $v_i$ is associated with a price $p_i$. For any two AS-vertices $v_i$ and $v_j$ that were projected by different I-vertices and $\mathsf{AS}(v_i)\cap \mathsf{AS}(v_j)\neq\emptyset$, there is an edge $e_{ij}$, called {\em AS-edge}, that connects $v_i$ and $v_j$. Each AS-edge is associated with a pair $(J, w_{i,j})$, where $J = \mathsf{AS}(v_i) \cap \mathsf{AS}(v_j)$, and the weight $w_{i,j}$ is the join informativeness of $D_i$ and $D_j$ on the join attributes $J$. 
\end{itemize}
\end{nameddefinition}

Definition \ref{def:jgraph} only specifies the weight of AS-edges. Next, we define the weight of I-edges. For any two I-vertices $v_i$ and $v_j$, let $AE$ be the set of AS-edges $\{(v_k, v_l)\}$ where $v_k$ ($v_l$, resp.) is an AS-vertex that $v_i$ ($v_j$, resp.) projects to. The weight of the I-edge $(v_i, v_j)$ is defined as the $w_{i,j} = \min_{(v_k, v_l)\in AE}w(k, l)$.

The I-layer can be constructed from the schema information of datasets in the marketplaces. Many existing marketplace platforms (e.g., \cite{azure,gwarehouse}) provide such schema information. AS-layer will be constructed from the data samples obtained from the marketplace.  Intuitively, there exists the trade-off between accuracy and cost. More samples lead to more accurate estimation of join informativeness, quality, and correlation at the AS-layer. But this costs \system for purchase of samples.

\nop{For any pair of AS-vertices $v_i$ and $v_j$, we define the distance of $v_i$ and $v_j$ as: \Boxiang{I am not sure if we need this. $d(v_i, v_j)$ seems not used in later sections}
\[d(v_i, v_j) = \min_{\Pi= v_i\dots v_j}w(\Pi).\]
}

Given a set of $n$ data instances $\mathcal{D}$ , its join graph contains $n$ instance vertices and $\Sigma_{i=1}^{n} (2^{m_i}-m_i-1)$ AS-vertices, where $m_i$ denotes the number of attributes in $D_i$. 
Figure \ref{fig:jgraph} shows an example of the join graph. 
An important property of the join graph is that all the AS-edges that connect the AS-vertices in the same instances with the same join attributes always have the same weight (i.e., the same join informativeness). For instance, the edges ($D_1. AB$, $D_2.BC$) and ($D_1. BC$, $D_2.BD$)  have the same weight. 
Formally,
\begin{property}
\label{property:equalweight}
For any two AS-edges $(v_i, v_j)$ and  $(v_i', v_j')$, if $v_i$ and $v_i'$ were associated with the same I-vertex, as well as for $v_j$ and $v_j'$, and $\mathsf{AS}(v_i)\cap \mathsf{AS}(v_j)= \mathsf{AS}(v_i')\cap \mathsf{AS}(v_j')$ (i.e., they have the same join attributes), then the two edges $(v_i, v_j)$ and  $(v_i', v_j')$ are associated with the same weight.
\end{property}
Property \ref{property:equalweight} is straightforward by following the definition of the join informativeness. For example, consider the join graph in Figure \ref{fig:jgraph}, the edge $(D_1.BC, D_2,BD)$ has the same weight as $(D_1.BC, D_2,BDE)$, as well as $(D_1.AB, D_2,BE)$ (not shown in Figure \ref{fig:jgraph}). 
Property \ref{property:equalweight} is in particular useful since it reduces the complexity of graph construction to exponential to the number of join attributes, instead of exponential to the number of all attributes. 
We will also make use of this property during graph search (Section \ref{sc:algo}) to speed up the search algorithm. 

We must note that the join informativeness does not have the {\em monotone} property (i.e., the join on a set of attributes $J$ has higher/lower join informativeness than any subset $J'\subset J$).  As shown in Figure \ref{fig:jgraph}, the join informativeness (i.e., weight) of join attributes $BC$ is higher than that of join attribute $B$, but lower than that of the join attribute $C$. 

Given a join graph, the source and target attributes $\mathcal{A_S}$ and $\mathcal{A_T}$ can be represented as special vertices in the join graph, formally defined below. 

\begin{table}
\begin{center}
\begin{tabular}{|c|c|}\hline
 Target attribute set & Covered instance vertex  \\\hline 
 $\{AB\}$&  $v_{1}$, $v_{2}$, $v_{3}$ (3 vertices)\\\hline
 $\{A\}$&  $v_{1}$, $v_{2}$, $v_{3}$, $v_{4}$ (4 vertices)\\\hline
 $\{B\}$&  $v_{1}$, $v_{2}$, $v_{3}$, $v_{5}$ (4 vertices)\\\hline
 $\{C\}$&  $v_{5}$, $v_{6}$ (2 vertices)\\\hline
 $\{BC\}$&  $v_{5}$, $v_{7}$ (2 vertices)\\\hline
\end{tabular}
\caption{\label{table:target} An example of target vertex sets}
\end{center}
\vspace{-0.3in}
\end{table}

\begin{nameddefinition}{Source/target vertex sets}
Given the source instances $\Set{S}$, the source attributes $\mathcal{A_S}$, and the target attributes $\mathcal{A_T}$, and a join graph $G$. An instance vertex $v\in G$ is a {\em source} I-vertex if its corresponding instance $D \in \Set{S}$. An instance vertex $v\in G$ is a {\em target} I-vertex if its corresponding instance $D$ is a source instance in $\Set{S}$ that contains at least one target attribute $A\in\mathcal{A_T}$. A set of AS-vertices $VS$ is a {\em source} ({\em target}, resp.) AS-vertex set if $\cup_{\forall v_i\in VS} \mathsf{AS}(v_i)$ = $\mathcal{A_S}$ ($\mathcal{A_T}$, resp.). In other words, the AS-vertex set covers all source (target, resp.) attributes. 
\end{nameddefinition}

Since some attributes may appear in multiple instances, there exist multiple source/target AS-vertex sets. Example \ref{exp:targetset} shows an example of the possible target vertex sets.

\begin{example}
\label{exp:targetset}
Consider a set of target attributes $\mathcal{A_T} = \{A, B, C\}$. These target attributes can be covered by four different options: 
\begin{itemize}
\item {\em Option 1}: they are covered by two instances, one containing \{A, B\} and the other containing \{C\}; 
\item {\em Option 2}: they are covered by three instances, one containing \{A\}, the second containing \{B\}, and the third containing \{C\}; 
\item {\em Option 3}: they are covered by two instances, one containing \{A\} and the other containing \{B, C\}; and
\item {\em Option 4}: they are covered by two instances, one containing \{A, B\} and the other containing \{B, C\}. 
\end{itemize}
Table \ref{table:target} shows all possible target instance vertices. Following Table \ref{table:target}, there are 3$\times 2 = 6$ possible target AS-vertex sets for Option 1, 4$\times$ 4 $\times$ 2 = 32 target AS-vertex sets for Option 2, 4$\times$ 2 = 6 sets for Option 3, and 3$\times$ 2 = 6 sets for Option 4. Note that some of these target vertex sets are duplicates, e.g., $v_{5}^{BC}$ covers both attribute sets \{C\} and \{B,C\}. In total there are 43 unique target AS-vertex sets. 
\end{example}

Based on the join graph and the source/target AS-vertex sets, the data acquisition problem is equivalent to finding a subgraph named {\em target graph} in the join graph, which corresponds to the instances for purchase. Next, we formally define the {\em target graph}. 
\begin{nameddefinition}{Target Graph}
Given a join graph $G$, a set of source AS-vertex sets $\mathcal{SV}$, 
and a set of target AS-vertex sets $\mathcal{TV}$, 
a connected sub-graph $TG\subseteq G$ is a {\em target graph}  if 
there exists a source AS-vertex set $SV\in \mathcal{SV}$ and a target AS-vertex set $TV\in \mathcal{TV}$ such that $TG$ contains both $SV$ and $TV$. 
\end{nameddefinition}

The target graph requires that it covers all source and target attributes. Next, we define the price, weight, and the quality of the target graph. Given a target graph $TG$, the price of $TG$ is defined as $p(TG) = \sum_{\forall_{v_i\in TG}}p_i$.
The {\em weight} of $TG$ is defined as: 
$w(TG) = \sum_{\forall (v_i, v_j)\in TG} w(i, j)$.
And the {\em quality} of $TG$, denoted as $Q(TG)$, is defined per Equation \ref{eqn:qmdmf}.  
Based on these definitions, now the data purchase problem can be mapped to the following graph search problem. 

\noindent{\bf Problem statement} Given a join graph $G$, a set of source AS-vertex sets $\mathcal{SV}$, a set of target AS-vertex sets $\mathcal{TV}$, and a budget $B$, find the {\em optimal target graph} (OTG) $G^*\subseteq G$ that satisfies the following constraint:
\begin{equation}
\label{eqn:of}
\begin{aligned}
& {\text{Maximize}}
& & CORR(\mathcal{A_S}, \mathcal{A_T})\\
& \text{Subject to}
&& w(G^*) \leq \alpha, \\
&&& Q(G^*) \geq \beta,\\
&&& p(G^*) \leq B,
\end{aligned}
\end{equation}
where $\alpha$ and $\beta$ are user-specified threshold for join informativeness and quality. 
We have the following theorem to show the complexity of the graph search problem. 
\begin{theorem}
\label{theorem:nphard}
The OTG search problem is NP-hard. 
\end{theorem}

The proof of Theorem \ref{theorem:nphard} can be found in our full paper \cite{cost2018li}. 

\vspace{-0.1in}
\section{Online Phase: Data Acquisition}
\label{sc:algo}

Given the acquisition request $(\mathcal{A_S}, \mathcal{A_T}$) with his source data $\Set{S}$, the brute-force approach is to search all possible target graphs in the join graph to find the best one (i.e., $\mathcal{A_S}$ and  $\mathcal{A_T}$ have the largest correlation). Obviously this approach is not scalable, given the fact that there is a large number of data instances in the data marketplace, and each data instance has exponentially many choices of attribute sets. Therefore, we design a heuristic algorithm based on Markov chain Monte Carlo (MCMC) method. 
The intuition is that with fixed $(\mathcal{A_S}, \mathcal{A_T})$, a large target AS-vertex set usually renders the join with small correlation and high join informativeness over longer join paths. Our algorithm consists of two steps. First, we find the minimal weighted I-layer graphs (I-graphs) (i.e.,  minimal join informativeness and thus higher correlation). The I-vertices in the subgraph correspond to the data instances. Second, we find the optimal target graph at the AS-layer (AS-graphs) from the minimal weighted I-graphs; the AS-vertices in the target graph correspond to the attribute sets of the instances identified by the first step. 
Next, we present the details of the two steps. 

\nop{
\nop{\begin{algorithm}
\label{alg:main}
\SetAlgoLined
     \SetKwInOut{Input}{Input}
     \SetKwInOut{Output}{Output}
 \Input{Join graph $G$, source instances $\Set{S}$, a set of source vertex sets $\mathcal{A_S}$, a set of target vertex sets $\mathcal{A_T}$.}
 \Output{A set of AS-vertices of join graph $G$}
 Initialize ITree = \{\}\;
 \{Step 1.\}\;
 \For{each $S \in \mathcal{S}$}{
  \For{each $T \in \mathcal{T}$}{
   Temp:= FindMinG\_ILayer(G, S, T)\{Algorithm 2\}\;
   \If{Temp has fewer nodes than ITreeSet}
      {ITreesSet:=Temp\;
      }  
   }
   }
   \{Step 2.\}\;   
   AttributeSetTree:= FindJoinTree\_AttSet(G, ITree) \{Algorithm 3.\}\;      Return AttributeSetTreeSet; 
\caption{Main function of search algorithm}
\end{algorithm}}

\begin{algorithm}
\label{alg:main}
\SetAlgoLined
     \SetKwInOut{Input}{Input}
     \SetKwInOut{Output}{Output}
 \Input{Join graph $G$, source instances $\Set{S}$, a set of source vertex sets $\mathcal{A_S}$, a set of target vertex sets $\mathcal{A_T}$.}
 \Output{A set of AS-vertices of join graph $G$}
 \{Step 1.Find the minimal target graph on instance layer\}\;    
    $T^*:=FindMinG\_ILayer(G',S,T)$\;
 \{Step 2. Find the minimal target graph on attribute set layer\}\;
    $AttributeSetTree:=FindJoinTree\_AttSet(G',T^*)$;
Return AttributeSetTree\;
\caption{Main function of search algorithm}
\end{algorithm}
}
\nop{
\subsubsection{Step 1: Construct Weights at the I-Layer of Join Graph}

The pseudo code is shown in Algorithm \ref{alg:wg}. 
\begin{algorithm}
\SetAlgoLined
     \SetKwInOut{Input}{Input}
     \SetKwInOut{Output}{Output}
     \Input{Join graph $G$}
     \Output{The weighted graph $G'$ corresponds to $G$}
\For{$e_{i,j}\in E$} {
	$X_{i,j}=D_i\cap D_j$\;
    $Temp=1$\;
    \For{$J\subset X_{i,j}$} {
        $JI(D_i, D_j, J)=\frac{H(D_i.J,D_j,J)-I(D_i.J,D_j.J)}{H(D_i.J,D_j,J)}$\;
        \If{$JI(D_i,D_j,J)<Temp$} {
        	$Temp=JI(D_i,D_j,J)$\;
            $J^*=J$
        }
    }
    $w_{i,j}=JI(D_i,D_j,J^*)$\;
}
Return $G'$
\label{alg:wg}
\caption{CreateWeightedILayer(): create the weights at the I-layer of the join graph}
\end{algorithm}
}

\subsection{Step 1: Find Minimal Weighted Graphs (I-graphs) at I-layer}
\label{sc:step1}
Given the source and target vertices, our graph search problem can be transformed to the classic Steiner tree problem that searches for a minimal tree that connects the source and target vertices. The Steiner tree problem is NP-hard. The complexity of the approximation algorithm \cite{vazirani2013approximation} is quadratic to the number of nodes, which may not be acceptable for large graphs. In this paper, we design a heuristic algorithm whose complexity is logarithmic to the number of nodes in the graph. 
Our algorithm extends the approximate shortest path search algorithm \cite{Gubichev:2010:FAE:1871437.1871503}. The key idea is to randomly pick a set of I-vertices as the {\em landmarks}. For each I-vertex in $G$, our algorithm pre-computes and stores the shortest weighted paths to these landmarks, by using the shortest path search algorithm in \cite{Gubichev:2010:FAE:1871437.1871503}. 
Then given the source and target AS-vertex sets $\mathcal{A_S}$ and $\mathcal{A_T}$, for each landmark $v_m$, the algorithm constructs a graph by connecting each vertex in $\mathcal{A_S}\cup \mathcal{A_T}$ and $v_m$, via their shortest weighted paths. The output I-graph is constructed as the union of all the shortest weighted paths.  
If the total weight of the I-graph exceeds $\alpha$, there does not exist a target graph that satisfies the join informativeness constraint. Therefore, the algorithm returns no output. Due to space limit, we omit the pseudo code. It can be found in our full paper \cite{cost2018li}.

\nop{
\begin{algorithm}
\SetAlgoLined
     \SetKwInOut{Input}{Input}
     \SetKwInOut{Output}{Output}
     \Input{Join graph $G$, a source vertex set $SV$, a target vertex set $TV$.}
 \Output{The minimal target graph that connects all vertices in $SV$ and $TV$.}
 Randomly pick a set of I-vertices from $G$ as the seeds \Wendy{why do we need the seeds? do you mean the landmark vertices?}\;
 Initialize output $T^*=\emptyset$\;
 \For{each seed $v_S$}
 {
  Initialize a graph $T'$, which includes $v_S$ and $SV\cup TV$\;
  \For{each vertex $v\in SV\cup TV$}{
    Find the weighted shortest path between $v_S$ and $v$ in $G$\ \Wendy{the citation of the algorithm that is used here? how to deal with weights?}; 
    Add edges and vertices of the path to $T'$\;  
    }
    \If{$T'$ is smaller than the trees in $T^*$}{
        $T^*$:= \{T'\};}
    \Else{
    \If{$T'$ has equal number of nodes as the trees in $T^*$}
  {Add $T'$ to $T^*$;}  
  }
  }
Return $T^*$; 
\label{alg:sp}
\caption{FindMinG\_ILayer(): find the minimal weighted target graphs at the instance layer}
\end{algorithm}
}

\nop{
\begin{algorithm}
\SetAlgoLined
     \SetKwInOut{Input}{Input}
     \SetKwInOut{Output}{Output}
     \Input{Weighted join graph $G$, a source vertex set $\mathcal{A_S}$, a target vertex set $\mathcal{A_T}$.}
 \Output{A set of minimal weighted graphs $\mathcal{IG}$, each connecting all vertices in $\mathcal{A_S}$ and $\mathcal{A_T}$.}
 Randomly pick a set of I-vertices in $G$ as the seeds\;
 Initialize output $\mathcal{IG}=\emptyset$\;
 \For{each seed $v_S$}
 {
  Initialize a graph $G'$ that includes $v_S$ and all I-vertices in $\mathcal{A_S}\cup \mathcal{A_T}$\;
  \For{each vertex $v\in \mathcal{A_S}\cup \mathcal{A_T}$}{
    Find the shortest weighted path \cite{Gubichev:2010:FAE:1871437.1871503} between $v_S$ and $v$ in $G$ at I-layer\; 
    Add edges and I-vertices of these paths to $G'$\;  
    }
    \If{the weight of $G'$ is smaller than the weights of the graphs in $\mathcal{IG}$}{
        $\mathcal{IG}$:= \{$G'$\};}
    \Else{
    \If{$G'$ has equal weight as the minimal weighted graphs in $\mathcal{IG}$}
  {Add $G'$ to $\mathcal{IG}$;}  
  }
  }
  \If{for any $G\in \mathcal{IG}$, the weight is greater than $\alpha$} {
  	$\mathcal{IG}=\emptyset$;
  }
  Return $\mathcal{IG}$; 
\caption{\label{alg:sp}FindMinG\_ILayer(): find the minimal weighted graphs at the I-layer}
\end{algorithm}
}

\subsection{Step 2: Find Optimal Target Graphs (AS-graphs) at AS-layer}
\label{sc:step2}
\nop{\begin{algorithm}
   \begin{minipage}{\hsize}
   \SetAlgoLined  
     \SetKwInOut{Input}{Input}
     \SetKwInOut{Output}{Output}
 \Input{A set of minimal trees $T^*$ at the instance layer}
\Output{A tree $\mathcal{T}$ at the AS-layer}
  \For{Each instance tree $T\in T^*$}{
  	  Initialize $i$:=0\;
      Initialize $Temp=\emptyset$\;
      Initialize MAX:=0\;
       \While{$i\leq\ell$}{
         Randomly pick an edge $e_{ij}$ from $T$ by follow a uniform distribution\;
         $J$: = common attributes of $v_i$ and $v_j$ \;
         Randomly select $J_0\subseteq J$\; 
         Accept $J_0$ by probability $min(1, \frac{Q(\pi_{J_0}(D))}{Q(\pi_{J}(D))})$\;
         \If{$J_0$ is accepted}{                   
           $J:=J_0$\;
           From $J$, construct the projection attributes $A_i$ and $A_j$ of $D_i$ and $D_j$ that $v_i$ and $v_j$ corresponds to\;
           Add $\pi_{A_i}(D_i)$ and $\pi_{A_j}(D_j)$ to $Temp$\;                    \If{$F(Temp)>MAX$ \{F(): Formula \ref{def:objfunction}\}}{
            $\mathcal{T}:=Temp$;}
          }
         i:=i+1 \;
       }
     }  
   Return $\mathcal{T}$\;
\label{alg:mcmc} 
\end{minipage}%
\caption{FindJoinTree\_AttSet(): find optimal target graphs at AS-layer}
\end{algorithm}}

\begin{algorithm}
\SetAlgoLined  
\SetKwInOut{Input}{Input}
\SetKwInOut{Output}{Output}
\Input{A minimal weighted I-graph $\mathcal{IG}$}
\Output{A target graph $G^*$ at the AS-layer}
	$G^*=NULL$\;
	$Max=0$\;
    $TG=\mathcal{IG}$\;
	\For{$i=1$ to $\ell$}{

        Randomly pick an edge $e_{i,j}\in TG$\;
        Randomly pick a different edge $e'_{i,j}$ between $(v_i,v_j)$\;
        Let $TG'$ be the new target graph\;
        \If{$p(TG')\leq B$ $\wedge$ $w(TG')\leq \alpha$ $\wedge$ $Q(TG')\geq \beta$} {
        \If{accept $e'_{i,j}$ by probability $\min(1,\frac{CORR(TG')}{CORR(TG)})$}{
        	$TG=TG'$\;
            \If{$CORR(TG)>Max$} {
				$G^*=TG$\;
				$Max:=CORR(G^*)$\;
			}
        }
	}{}
    }
Return $G^*$\;
\caption{\label{alg:mcmc} FindJoinTree\_AttSet(): find optimal target graph at attribute set layer}
\end{algorithm}

\nop{
\begin{algorithm}
   \SetAlgoLined  
     \SetKwInOut{Input}{Input}
     \SetKwInOut{Output}{Output}
\Input{A set of minimal weighted I-graph $\mathcal{IG}$}
\Output{A target graph $G^*$ at the AS-layer}
  \For{each I-graph $IG\in\mathcal{IG}$}{
      Initialize $i$:=0\;
      Initialize $Max:=0$\;
       \While{$i\leq\ell$}
        {
        \If{$Satisfy(IG)$} {
        	\If{$Corr(IG)>Max$} {
            	$G^*=IG$\;
                $Max:=Corr(G^*)$\;
            }
        }
        \{Pick the edge whose join attribute is to be replaced\}\;
        Initialize edge $e=NULL$\;
        \If{$p(IG)>B$}
        {
         $e(v_i, v_j)$:= pick edge by the price-based transition probability \Wendy{1. add def. of edge price. 2. add def. of transition probability. higher price has higher transition probability}\;
        }
        \If{$w(IG)>\alpha$}
        {
         $e(v_i, v_j)$: = pick edge by the weight-based transition probability\; \Wendy{add def. of transition probability.}      
        }
        \If{$q(IG)>\beta$}
        {
          $e(v_i, v_j)$: = randomly pick an edge from $IG$.
        }
         $i:=i+1$\;
       }     
     }  
   Return $G^*$\;
\label{alg:mcmc} 
\caption{FindJoinTree\_AttSet(): return the optimal target graph at the  AS-layer}
\end{algorithm}
}

\nop{
\begin{algorithm}
   \begin{minipage}{\hsize}
  \SetAlgoLined
     \SetKwInOut{Input}{Input}
     \SetKwInOut{Output}{Output}
 \Input{A tree $T_S=(V_S,E_S)$ on instance layer calculated by Algorithm 2, and a set of join attributes $\mathcal{C}$}
\Output{A tree $T=(V, E)$ on attribute set layer}
	$V=\{\}$\;
    $E=\{\}$\;
       \For{$J_{i,j}\in \mathcal{C}$}{
			\If{$V_i$ is $D_S$} {
            	$V = V + D_S$
            }
            \If{$V_j$ contains target attribute(s) $t$} {
            	$V = V + J_{i,j}\cup t$
            }
            \Else{
            	$V = V + \cup_{e_{i,j_k}\in E_S} J_{i, j_k}\cup_{e{j',i}\in E_S} J_{j', i}$
            }
       }
       \For{$e_{i,j} \in E_S$} {
       		$E = E + e_{v_i, v_j}$, where $v_i, v_j \in V$
       }
       Return $T$
\label{alg:mcmc} 
\end{minipage}%
\caption{ConstructTree(): build the tree on attribute set layer}
\end{algorithm}
}

Based on the data instances selected by Step 1, Step 2 further selects the projection attributes of these instances by searching at the AS-layer of the minimal weighted I-graph. Algorithm \ref{alg:mcmc} shows the pseudo code of Step 2. 
The key idea of Step 2 is to generate a sample of the graph at AS-layer iteratively by replacing the join attribute set of one edge $e_{i,j}$ with a different join attribute set. The algorithm runs $\ell$ iterations and keeps the sample of the largest correlation between the source and target vertices (Line 11 - 14). 
Some samples may not satisfy the constraints on weights, quality, and/or price.
For each new sample, we first check if it satisfies these constraints (Line 8). 
After that, we use a Markov Chain Monte Carlo (MCMC) process to generate samples with high correlation, so as to maximize the utility of the output. 
In particular, for each graph, the algorithm randomly picks an edge $e_{i,j}$  (Line 5), which represents the join between two instances $D_i$ and $D_j$. From all the possible join attributes between $D_i$ and $D_j$, the algorithm randomly picks one, which corresponds to the AS-edge $e'_{i,j}$, and replaces $e_{i,j}$ with $e'_{i,j}$ by the acceptance probability $\min(1,\frac{CORR(TG')}{CORR(TG)})$ (Line 9). Intuitively, the target graph of high correlation is accepted with high probability. During the iterations, the algorithm always keeps the graph sample that has the largest correlation between the source and target vertices, and return it after $\ell$ iterations.

\nop{
Step 2 selects the projection attributes for each of these instances. 
Thus, the goal of Step 2 is to search for a set of AS-vertices at the AS-layer of the join graph, one vertex for each instance vertex in $V$. The naive method is to try all attribute sets for each data instance. There are  $\Pi{i=1}^{k} (2^{m_i} - 1)$ candidates in the search space, where $m_i$ is the number of attribute of the instance that $v_i\in V$ has. To avoid the expensive search, we design a heuristic Markov chain Monte Carlo (MCMC) method. Instead of searching through all possible attribute sets of the instances, we set the search space as the set of all possible {\em join attribute sets} of each pair of instances, and use these join attribute sets to decide the projection attributes of the instances.  Therefore, our algorithm consists of two steps: (1) find the join attributes between instance pairs; (2) construct the projection attributes for each instance from the join attributes. 

\noindent{\bf Step 2.1: find join attributes between instance pairs.} 
Note that the minimal target graphs returned by Step 1 indeed indicate how these instances will be joined. Intuitively, given two instances $D_i$ and $D_j$ that join on $k$ attributes (e.g., \{A, B\}), their projection instances can join on possible $2^{k}-1$ attributes (e.g., \{A\}, \{B\}, and \{A, B\}). Which join attributes to pick is decided by the data quality of the corresponding projection instances. For example, we prefer to pick  $\pi_{\{A\}}(D_i)$ if its data quality is better than $\pi_{\{A, B\}}(D_i)$ and $\pi_{\{A, B\}}(D_i)$.  
Based on this reasoning, we model the graph search algorithm at the attribute set level as a random walk through a Markov chain, where the transition probability moving from an attribute set vertex $v_i$ to $v_j$ is calculated as: 
\[prob(v_i\rightarrow v_j) = min(1, \frac{Q(\pi_{A_j}(D))}{Q(\pi_{A_i}(D))}),\]
where $A_i$ and $A_j$ are the attributes that $v_i$ and $v_j$ correspond to. We design the algorithm as following. 

For each pair $(v_i, v_j)$ where $v_i, v_j$ are picked by Step 1, let $J$ be the join attributes of the data instances $D_i$ and $D_j$ that $v_i$ and $v_j$ correspond to. Our algorithm first chooses an arbitrary attribute set $J_0\subseteq J$ to be the first sample. Then the algorithm walks through the paths of the join graph at the attribute set layer for $\ell$ iterations, where $\ell$ is a user-specified parameter. In each iteration, the algorithm follows a uniform distribution to pick a candidate $J$ for the next sample of join attribute set, given the previous sample value $J_0$. Then the algorithm follows the transition probability to accept or reject the new sample. Intuitively, the samples (i.e., attribute sets) of higher quality will be picked with higher transition probability. Algorithm \ref{alg:mcmc} (Line 7 - 11) shows the pseudo code of the MCMC approach. 
}

\nop{
\noindent{\bf Step 2.1: Revise projection attributes at AS-edges.}  
We use $\mathsf{AS}(v)$ to denote the attribute set that the vertex $v$ corresponds to. 
Consider an I-graph $\mathtt{IG}$, with each I-edge $(v_i, v_j)$ is associated with a pair ($w_{i,j}$, $J$), where $w_{i,j}$ is the weight of the AS-edge $(v_k, v_l)$ ($v_k$ and $v_l$ are projected by $v_i$ and $v_j$ respectively), and $J = \mathsf{AS}(v_k) \cap \mathsf{AS}(v_l)$. Following the definition of I-edge weight, apparently, $w_{i,j}$ is the best join informativeness of the join of all possible subsets of data instances that $v_i$ and $v_j$ correspond to.  We revise $\mathtt{IG}$ based on how it violates the constraint. 
\begin{itemize}
\item If $\mathtt{IG}$ fails the weight constraint (i.e., $w_{\mathtt{IG}}\geq\alpha$), then \Wendy{???}.
\item If $\mathtt{IG}$ fails the quality constraint (i.e., $q_{\mathtt{IG}}\leq\beta$), then \Wendy{???}.
\item If $\mathtt{IG}$ fails the price constraint (i.e., $w_{\mathtt{IG}}\geq B$), then \Wendy{???}.
\end{itemize}

\noindent{\bf Step 2.2: Construct projection attributes from join attributes.} Step 2.1 determines the join attributes of each instance pair. By Step 2.2., the algorithm constructs the projection attributes from the join attributes by Step 2.1. Specifically, 
for any three adjacent vertices $(v_i, v_j, v_k)$ in the minimal tree (by Step 1), without losing generality, we assume that their corresponding instances join by the following: $D_i$ joins with $D_j$ and $D_j$ joins with $D_k$. Then for the join attributes $J_{ij}$ (of $D_i$ and $D_j$) and $J_{jk}$ (of $D_i$ and $D_j$) that are decided by Step 2.1., the algorithm picks $J_{ij} \cup J_{jk}$ as the projection attribute set of the instance $D_j$. For the first (last, resp.) instance of the join, its projection attribute set is constructed by union of the source (target, resp.) attributes and the join attribute. 
Based on the constructed projection attributes, the algorithm calculates the value of the objective function of the projection instances, and keeps the one of the best objective function. 
}

\subsection{Complexity Analysis}
\nop{Step 1 requires $[lg(n)]$ seeds, where $n$ is the number of the vertices at the I-layer. }
The complexity of Step 1 is $O(k(|\mathcal{A_S}|+|\mathcal{A_T}|)\log_2n)$, where $k$ is the average size of the minimal weighted graphs at the I-layer, and $n$ is the number of data samples on \system (i.e., the number of vertices at the I-layer of the join graph). In the experiments, we observe that $k$ is much smaller than $n$, especially when $n$ is large. The complexity of Step 2 is 
$O(\ell C_J)$, 
where $\ell$ is the number of iterations, and $C_J$ is the average join cost of the instances in the I-graph returned by Step 1. Note that $C_J$ depends on both the size of I-graphs, and the size of each instance in the I-graph. The total complexity of the graph search algorithm is dominated by Step 2, which is $O(\ell C_J)$. 

\nop{
\subsubsection{Optimization} 

\noindent{\bf Triangle Inequality.}
We can further optimize the MCMC graph search method. One possible optimization is to utilize the triangle inequality on join strength during the search.

\begin{property}
\label{property:ti}
Given two edges $(v_i, v_j), (v_j, v_k)\in G$, it holds that 
\[JS(\{S_i, S_j, S_k\})\leq min(JS(\{S_i, S_j\}), JS(\{S_j, S_k\})),\]
where $D_i$ is the data instance that $v_i$ corresponds to.
\end{property}

Property \ref{property:ti} shows that the search algorithm cannot be simply optimized by  considering the minimum of edge weights of all trees. Instead, we optimize the search space with the {\em monotone join strength} property of the join graph. 

\noindent{\bf Estimation of join strength.} Join can be very expensive. A possible solution is to adapt the existing algorithms \cite{haas1993fixed,estan2006end,Vengerov:2015:JSE:2824032.2824051} to estimate the join size  quickly with high accuracy, instead of executing joins physically. In particular, we apply {\em correlated sampling} approach \cite{Vengerov:2015:JSE:2824032.2824051} to estimate the join size. The basic idea of the algorithm is to pick a sample of each data instance, and join these samples. The true join size is estimated by scaling the join size of the samples. Join strength then can be computed based on the estimated join size. 

\Wendy{Seems it is impossible to estimate the data quality.}
{\bf Estimation of data quality of join results.} Given a set of joinable data instances $\mathcal{D}=\{D_1, D_2, \dots, D_k\}$, we have $\mathcal{C}=\{C(D_i, \mathcal{F}_i)|1\le i \le k\}$. Then the quality of the join result can be calculated by the following steps:
\begin{itemize}
\item For each $C\in \mathcal{C}$, treat it as the {\em correct subtable} of each data instance, let $\mathcal{CJ}$ = $\bowtie^k_1 C_i$ be the join result of these correct subtables
\item Let $\mathcal{J}$ be the join result of $\mathcal{D}$
\item Then $Q(\mathcal{J})$ is $\frac{\|\mathcal{CJ}|}{|\mathcal{C}|}$
\end{itemize}
The sizes of both $\mathcal{CJ}$ and $\mathcal{J}$ can be estimated in the way we have described easily, thus the estimation of $Q(\mathcal{J})$ can be calculated. In this way, the estimation of quality can be done by using join size estimation. 

\Wendy{1. The original algorithm is wrong. Change algorithm. 2. Add an example to explain. 3. Add complexity of the two estimation algorithms.}

For estimation of the join size of two datasets, assume the size of one dataset is $n$, another is $m$. It first requires one-pass through both datasets to generate the samples, the time complexity of which is $O(n)$ and $O(m)$ respectively. Assume the sample sizes are $n'$ and $m'$, where $n'<<n$ and $m'<<m$. Then the actual join is done on the samples, and the time complexity is $O(n'm')$(?). After getting the join result of the samples, the actual join size is calculated by scaling the size of the joined samples, which is $O(1)$. The time complexity of estimating the join size of multiple tables can be analyzed in a similar way.
}

\nop{
dynamic programming and can reach global optimality, which is shown as Algorithm \ref{algo:ex}. This algorithm works in a similar way as Floyd-Warshall algorithm that finds all-pairs shortest-paths. It computes an $n \times n$ matrix, where $n$ is the number of 2-attribute sets of all tables from this database. And each entry $(i, j)$ stores the maximal objective value of a join path from $i$ to $j$, which serves as the {\em distance} in the original algorithm. When $i$ contains source and $j$ contains destination, the value at $(i, j)$ is a candidate of the maximal objective value from source to destination, and the maximal one among all these values is the maximal objective value from source to destination, and the corresponding join path is the {\em best} path we want to find. Like Floyd-Warshall, the algorithm computes the 1-hop objective value matrix at first, then computes the 2-hop objective value matrix based on the 1-hop matrix, keeps computing the objective value matrix until gets the $n$-hop objective value matrix, in which we can find the maximal objective value from source to destination and the best join path.
\begin{algorithm}
\label{algo:ex}
\caption{Exact Algorithm}
  \begin{algorithmic}[1]
	\REQUIRE A source attribute set $S$, a destination attribute set $D$, and all the 2-attribute sets of all tables from the database \{$S_1$, $S_2$, ..., $S_n$\}; 
	\ENSURE Find the join path of optimal objective value from $S$ to $D$. 
	\STATE{Let $m^{(k)}_{ij}$ be the objective value of any pair of vertices $S_i$ and $S_j$ where the inner vertices along the join path are from \{$S_1$, $S_2$, ... $S_k$\}}
	\STATE{Let $M^{(k)}$ be the $n\times n$ $k$-hop objective value matrix}
	\STATE{$M^{(0)}$ stores the objective values of all joinable pairs of vertices; for those who are not joinable, the value is $\infty$}
	\FOR{$k=1$ to $n$}
		\STATE{Let $M^{(k)}=(m^{(k)}_{ij})$ be a new $n\times n$ matrix}
		\FOR{$i=1$ to $n$}
			\FOR{$j=1$ to $n$}
				\STATE{Let $P_{old}$ be the path from $S_i$ to $S_j$ in $M^{(k-1)}$}
				\STATE{$P_{new}=P_{old}\union S_k$}
				\STATE{$temp=F(P_{new})$}
				\STATE{$m^{(k)}_{ij}=$max$(m^{(k-1)}_{ij}, temp)$}
			\ENDFOR
		\ENDFOR
	\ENDFOR
	\STATE{$maxobj=0$}
	\FOR{each $m^{(n)}_{ij}$ in $M^{(n)}$}
		\IF{$S_i$ contains source and $S_j$ contains destination}
			\IF{$m^{(n)}_{ij}>maxobj$}
				\STATE{$maxobj=m^{(n)}_{ij}$}
			\ENDIF
		\ENDIF
	\ENDFOR
	\STATE{Return $maxobj$}
\end{algorithmic}
\end{algorithm}
}
\nop{
\begin{algorithm}
\caption{Local Optimality}
  \begin{algorithmic}[1]
	\REQUIRE A source attribute set $S$, a destination attribute set $D$, and a set of candidate datasets \{$D_1, D_2, ..., D_N$\}, the correspoding join graph $G$; 
	\ENSURE Find the path of optimal objective value from $S$ to $D$. 
	\STATE{{\bf Step 1}: Find the 1-hop 2-attribute sets of $S$}
		\FOR{all $D_i$}
			\FOR{all joinable 2-attribute sets $S_i$}
				\STATE{Measure the quality of $S_i$}
			\ENDFOR
			\FOR{all different join attribute $J_i$}
				\FOR{each set of $S_i$ that join with $S$ on the same join attribute $J_i$}
					\STATE{find the $S_i$ with the best quality}
					\STATE{Measure the objective value of that $S_i$}
				\ENDFOR
			\ENDFOR
		\ENDFOR
		\STATE{Pick the $S_i$ of maximal objective value}
	\STATE{{\bf Step 2}: }
		\STATE{Remove all the unpicked 1-hop 2-attribute sets from $G$}
		\STATE{Use $S_i/J_i$ as the new seed, repeat Step 1 to pick the best 1-hop 2-attribute set}
	\STATE{Repeat Step 2, until reach $D$}
\end{algorithmic}
\end{algorithm}
}

\nop{
We have a heuristic algorithm that can reach global optimality, which is shown in Algorithm \ref{algo:ske}. This approach is based on the {\em sketch-based} shortest path estimation for large graphs \cite{Gubichev:2010:FAE:1871437.1871503}. However, one of the fundamentals of this kind of methods is the {\em triangle-inequality} of distance functions, that is, for any distance function dist(), $s, d, l$ are three different nodes in the graph, we have \[\text{dist}(s, d)\le \text{dist}(s, l)+\text{dist}(l, d)\]. Unfortunately, in our join graph, the edge weight does not satisfy the requirements of a distance function, thus there is no triangle-inequality in our join graph.
\begin{example}
Given three nodes $S_1$, $S_2$, $S_3$, assume there is a path $P_{13}$ directly from $S_1$ to $S_3$, and the join result of this path is $R_{13}$; and another path $P_{123}$ from $S_1$ to $S_3$ through $S_2$, and the join result of this path is $R_{123}$. Then according to the definition of join strength of a path, $JS_{P_{13}}=\frac{|R_{13}|}{|S_1||S_3|}$, $JS_{P_{123}}=\frac{|R_{123}|}{|S_1||S_2||S_3|}$. And because of the commutability of join, $R_{123}$ equals to the join result of $R_{13}$ and $S_2$, so we have $|R_{123}|\le |R_{13}||S_2|$, $JS_{P_{123}}=\frac{|R_{123}|}{|S_1||S_2||S_3|}\le \frac{|R_{13}||S_2|}{|S_1||S_2||S_3|}=\frac{|R_{13}|}{|S_1||S_3|}=JS_{P_{13}}$, and it is not likely to have equal. 
For path quality, according to the definition, we have $error_{P_{13}}=\frac{|ER_{13}|}{|R_{13}|}$, $error_{P_{123}}=\frac{|ER_{123}|}{|R_{123}|}=\frac{ER_{13}\text{join}ER_2}{R_{13}\text{join}S_2}$, so there is no guarantee that which one will be bigger. In conclusion, there is no triangle-inequality for our objective function.
\end{example}

Because there is no triangle-inequality of the objective value, we cannot apply the sketch-based shortest paths estimation directly. To solve the problem, we need to notice that the join strength decreases sharply with the increase of number of hops, {\em i.e.} number of nodes along the join path, so naturally maximizing the join strength leads to minimization of the number of hops. Therefore, in this algorithm, we consider join strength and quality separately, and solve the problem in two steps. In the first step, we find the shortest paths in number of hops, which is also the number of supernodes as for next node we will never choose one from the same lattice, and this step maximizes the join strength. In the second step, we find the path with the best quality among all the shortest paths found in step 1. In this way, the final join path we found is expected to have the maximum objective value.

After finding the shortest paths on the supernode level, there may be multiple shortest paths on the node level. The following lemma shows that the number of shortest paths is bounded.
\begin{lemma}
Given a source attribute $S$, a destination attribute $T$, and a set  of candidate datasets $\{D_1, \dots, D_N\}$, where the number of attributes in each dataset is $m_1,\dots, m_N$, the maximum amount of the shortest paths that connect $S$ to $T$ is $N!\prod_{i=1}^N (m_i-1)$.
\end{lemma}
\begin{proof}
First, let $\phi(.)$ be a random permutation function for integers $\{1, \dots, N\}$. Following each permutation, there is a sequence of datasets $(D_{\phi^{-1}(1)}, \dots, D_{\phi^{-1}(N)})$ to connect $S$ and $T$. It is easy to see that there exist $N!$ permutations.

According to Theorem \ref{th:2attribute}, for each dataset, we always pick 2-attribute set as a hop in the join sequence. Let $\mathcal{A}_i$ be the 2-attribute set picked from the dataset $D_i$. To build a join path based on the permutation $\phi(.)$, it must be true that $|\mathcal{A}_{\phi^{-1}(i)}\cap \mathcal{A}_{\phi^{-1}(i+1)}|$ for any $1\leq i < N$. For each permutation, there are at most $\prod_{i=1}^N (m_i-1)$ 2-attributes to establish the join. 

Therefore, the maximum amount of the shortest paths that join $S$ and $T$ is $N!\prod_{i=1}^N (m_i-1)$.
\end{proof}
\begin{algorithm}
\label{algo:ske}
\caption{Sketch-based Approach}
  \begin{algorithmic}[1]
	\REQUIRE A source attribute set $S$, a destination attribute set $D$, and a set of candidate datasets \{$D_1, D_2, ..., D_N$\}, the correspoding join graph $G$; 
	\ENSURE Find the path of optimal objective value from $S$ to $D$. 
	\STATE{Find the shortest paths(in term of hops) from $S$ to $D$ using the sketch method \cite{Gubichev:2010:FAE:1871437.1871503}}
	\STATE{Among all the shortest paths, pick the path of the best value of objective function based on the estimated join strength and quality \cite{Vengerov:2015:JSE:2824032.2824051}} 
\end{algorithmic}
\end{algorithm}

\Wendy{1. Complexity of sketch-based approach? 2. Line 2 should change to pick the path of the best value of objective function. 3. Any optimization on calculating the join strength }

\Wendy{Complexity of sketch-based approach is $O(lgn)$? }

For the first step in Algorithm 4, the time complexity is $O(\log n)$, where $n$ is the number of nodes of the join graph. For the second step, its time complexity has already be dicuseed in Section 5.1.

Change and add of definitions:

superedge: if two super vertex are joinable, i.e. the two datasets are joinable, there is a superedge connecting two super vertices

$X_{i,j}$: given two datasets $D_i$ and $D_j$, $X_{i,j}=D_i\cap D_j$

$J_{i,j}$: a subset of $X_{i,j}$, $J_{i,j}\subset X_{i,j}$

$\mathcal{C}_{i,j}$: given two datasets $D_i$ and $D_j$, the set of all possible $J_{i,j}$ \Yanying{The notion is not good, but I cannot come up with anything better now, we can change it later}

because the output is a tree, for a given super vertex/vertex, the edge from its parent to the vertex is called the incident edge, the edges from the vertex to its children are called output edges

assumption: given two regular vertices, they must join on the maximal join attribute following natural join; given two datasets $D_i$ and $D_j$, they can join on any $J_{i,j}\subset X_{i,j}=D_i\cap D_j$; i.e, given two vertex, the join result is unique, given two super vertex, there are multiple possible different join result
}


\nop{
\section{Integration with Existing Data Marketplace Platforms}
\Wendy{ Concerns: (1) pricing function inconsistent with the existing data marketplace; (2) experiments to justify that using our middleware indeed can save money for the users; (3) change the architecture in Figure 1. }

\subsection{Entropy-based Pricing Function}
\label{sc:pf}

In this section, we present our entropy-based pricing function and its arbitrage-free property. 

Given an instance $D$ and a set of attributes $A$ of $D$, a {\em pricing function} $p()$ assigns to $\pi_{A}(D)$ ($\pi$: projection operator) a price in the domain $\mathbb{R}$. One of the issues is how to design the price function $p()$. A notion central to the pricing function is {\em arbitrage}. Informally, if query $\mathsf{Q}_1$ discloses more information than query $\mathsf{Q}_2$, it must hold that $p(\mathsf{Q}_1(D)) \geq p(\mathsf{Q}_2(D))$. Otherwise, the buyer has an arbitrage opportunity to purchase the desired information at a lower price. 
Formally, a pricing function is arbitrage-free it it satisfies the following requirement.
\begin{definition}
\label{df:arbitrage-free}
A data pricing function $p()$ is arbitrage-free if any pair of queries $\mathsf{Q}_1$ and $\mathsf{Q}_2$, if $D\vdash \mathsf{Q}_2 \rightarrow \mathsf{Q}_1$, then $p(\mathsf{Q}_2(D))\geq p(\mathsf{Q}_1(D))$.
\end{definition}

Next, we have the following important theorem to define an arbitrage-free pricing function for projection queries. We first define the {\em partition-based information entropy}.

\begin{nameddefinition}{Partition-based information entropy}
Given a data instance $D$ and an attribute set $X$, let $\pi_X=\{eq_X^1, \dots. eq_X^q\}$ be the set of equivalence classes of $D$ on the attribute set $X$. We define the {\em partition-based information entropy} of $X$, denoted as $H(X)$, that measures the minimum number of binary bits to represent the partitions on $X$. Formally,
\begin{equation}
\label{eq:entropy}
H(X)=-\sum_{\forall eq_X^i\in \pi_X} P(eq_X^i)\log_2 P(eq_X^i),
\end{equation}
where $P(eq_X^i)=\frac{|eq_X^i|}{n}$, and $|eq_X^i|$ is the size of the equivalence class $eq_X^i$.
\end{nameddefinition}

Given a pricing function $p()$, we say $p()$ is non-decreasing on the partition-based entropy $H(X)$ if for any two sets $X_1, X_2\subseteq\mathcal{A}$, if $H(X_1)< H(X_2)$, then $p(X_1) \leq p(X_2)$. Then we have the following theorem. 

\begin{theorem}
\label{theorem:entropy_af}
Given any data instance $D$ that contains FD $F: X\rightarrow Y$, for any pricing function $p()$ that is not decreasing on the partition-based information entropy $H(X)$, $p()$ is arbitrage-free.  
\end{theorem}

The proof of Theorem \ref{theorem:entropy_af} is included in Appendix. 

A straightforward way to define $p()$ is to define a linear function on the partition-based information entropy: 
\[p(D) = aH(D) + b,\]
where $a$ and $b$ are constants, in particular $a$ is a positive constant. Obviously, such function is not decreasing. In this paper, we use $p(D) = H(D)$. 

Finally, we define the price of a given set of instances. 
\begin{nameddefinition}{Price of a set of instances}
\label{def:costmtable}
Given a set of data instances $\mathcal{R}$, the total price of $\mathcal{R}$ is defined as the sum of the prices of individual prices.
\[p(\mathcal{R}) =\sum_{\forall R_i\in\mathcal{R}}p(R_i).\] 
\end{nameddefinition}

}

\begin{table*}
\begin{center}
\begin{tabular}{@{}|@{}c@{}|@{}c@{}|@{}c@{}|@{}c@{}|@{}c@{}|@{}c@{}|@{}c@{}|@{}}\hline
        & \# of instances &  Min. instance size & Max. instance size & Min. \# of attributes &   max. \# of attributes  & Avg \# of FDs per table\\
        & &  (\# of records) &  (\# of records) &  &   & \\\hline 
TPC-H & 8 & 5 (Region) &  6,000,000 (Lineitem) &  4 (Region) & 20 (Lineitem)  & 39 \\\hline  
TPC-E & 29 & 4 (Exchange)  & 10,001,048 (Watchitem) & 3 (Sector) & 28 (Customer)  & 33 \\\hline  
 \end{tabular} 
 \vspace{-.1in}
\caption{\label{table:dataset} Dataset description}
\end{center}
\vspace{-0.1in}
\end{table*}

\begin{figure*}[!htpb]
\centering
\begin{tabular}{ccc}
	\includegraphics[width=0.33\textwidth]{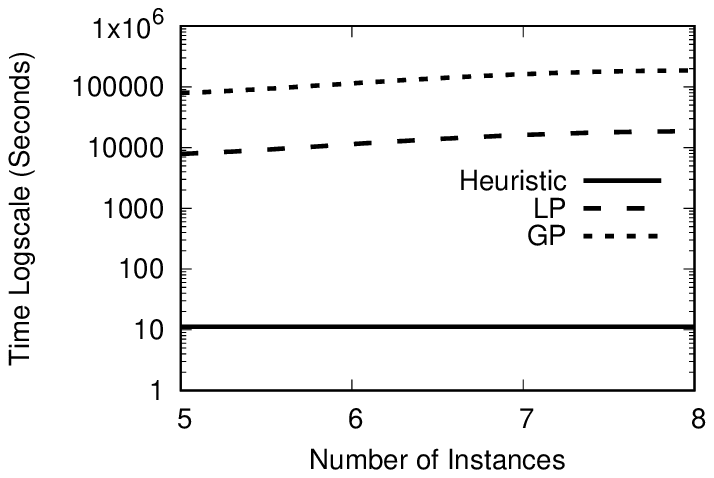}
    &
    \includegraphics[width=0.33\textwidth]{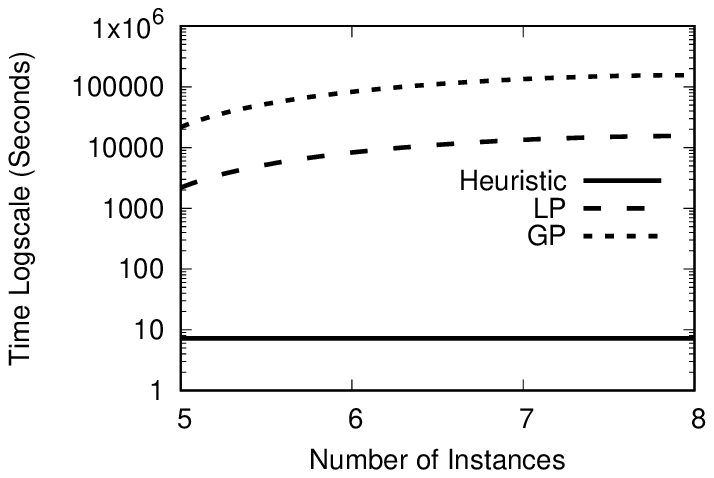}
    &
     \includegraphics[width=0.33\textwidth]{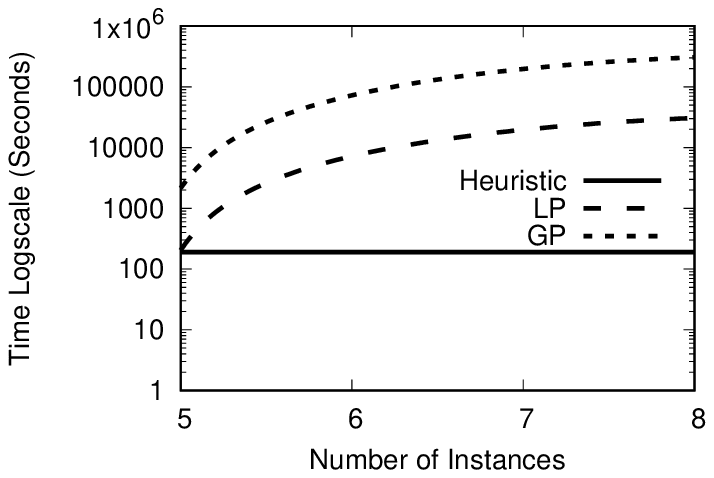}
    \\
    {\small (a) $Q_1$}
    &
    {\small (b) $Q_2$}
    &
    {\small (c) $Q_3$}
\end{tabular}
\vspace{-.1in}
 	\caption{\small \label{fig:timeperformance-numinstance} Time Performance w.r.t. various \# of instances and join length (TPC-H dataset)} 
\vspace{-0.1in}
\end{figure*}

\begin{figure*}[!htbp]
\centering
\begin{subfigure}[b]{0.3\textwidth}
	\centering
	\includegraphics[width=\textwidth]{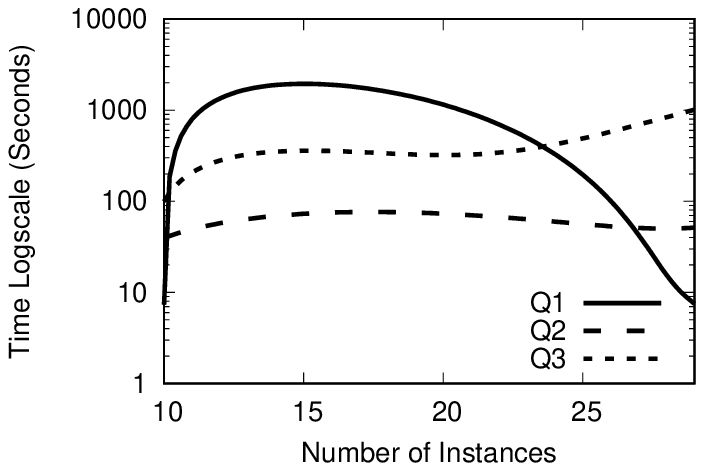}
    \caption{{\small Various \# of instances}}
\end{subfigure}
~
\begin{subfigure}[b]{0.3\textwidth}
	\centering
    \begin{tabular}{|c|c|c|c|}
    \hline
    \# of Instances & $Q_1$ & $Q_2$ & $Q_3$ \\\hline
    10 & 3 & 5 & 5 \\\hline
    15 & 7 & 4 & 7 \\\hline
    20 & 5 & 4 & 4 \\\hline
    25 & 5 & 4 & 6 \\\hline
    29 & 3 & 5 & 8 \\\hline
    \end{tabular}
    \vspace{0.2in}
    \caption{\small I-graph size }
\end{subfigure} 
~
\begin{subfigure}[b]{0.3\textwidth}
	\centering
    \begin{tabular}{|c|c|c|c|}
    \hline
    Budget Ratio & $Q_1$ & $Q_2$ & $Q_3$ \\\hline
    0.04 & 3.067 & N/A & N/A \\\hline
    0.06 & 6.487 & 30.62 & N/A \\\hline
    0.08 & 7.478 & 49.23 & 766.8 \\\hline
    0.10 & 7.478 & 51.43 & 980.9 \\\hline
    0.12 & 7.478 & 51.43 & 1009.8 \\\hline
    \end{tabular}
	\caption{\small Various budget ratio (N/A: not affordable)}
\end{subfigure}'
\caption{\small \label{fig:timeperformance-numinstance-e} Time performance w.r.t. various \# of instances and budget ratios (TPC-E dataset)}
\end{figure*}

\begin{table*}[h!]
\begin{center}
\begin{tabular}{|c|c|c|c|c|c|}\hline
   Query & Approach &  Correlation & Quality & Join Informativeness & Price \\\hline 
   \multirow{2}{*}{$Q_1$} & With \system & 12.51  & 0.07688 & 0.8906 & 60.65\\\cline{2-6} 
   & Purchase from data marketplace & 18.18  & 0.3819 & 0.8268 & 63.12 \\\hline 
   \multirow{2}{*}{$Q_2$} & With \system & 7.216 & 0.3968 & 1.248 & 59.78\\\cline{2-6}
   & Purchase from data marketplace & 7.965 & 0.4244 & 1.040 & 103.2 \\\hline
   \multirow{2}{*}{$Q_3$} & With \system & 3.609 &  0.08632 & 2.063 & 100.5 \\\cline{2-6}   
   & Purchase from data marketplace & 4.592 & 0.1026 & 2.104 & 106.4 \\\hline       
 \end{tabular}
\caption{\label{table:results} Comparison between data acquisition with \system  and purchase from the data marketplace directly (TPC-H dataset, budget ratio=0.13)}
\end{center}
\vspace{-0.1in}
\end{table*}

\section{Experiments}
\label{sc:exp}



\vspace{-0.05in}
\subsection{Experimental Setup}

\noindent{\bf Implementation \& Testbed.} We implement the algorithms in {\em Python}.  All the experiments are executed on a machine with 2 $\times$ Intel(R) Xeon(R) Silver 4116 CPU @ 2.10GHz, 12 cores, 24 processors, and 128 GB memory.

\noindent{\bf Datasets} We use the TPC-E\footnote{http://www.tpc.org/tpce/} and TPC-H\footnote{http://www.tpc.org/tpch/} benchmark datasets in our experiments. The details of these two datasets are shown in Table \ref{table:dataset}. The tables that are of the minimum/maximum number of records are shown in a pair of parentheses. The longest join path in TPC-H and TPC-E datasets are of length 7 and 8 respectively.  

\noindent{\bf Functional Dependency (FD)} We measure the number of FDs on the TPC-E and TPC-H datasets. We use $\theta=0.1$ as the as the user-defined threshold for FDS (i.e. the amount of records that do not satisfy FDs is less than 10\% of the dataset). We find that the number of FDs varies for each instance of TPC-H and TPC-E dataset. For example, there are 114 FDs in the {\em LineItem} table and 6 FDs in {\em Region} table in the TPC-H dataset. 

\noindent{\bf Data quality.} We modified 30\% of records of 6 tables in TPC-H dataset (except {\em region} and {\em Nation} tables), and 20 out of 29 tables in TPC-E dataset to introduce inconsistency into the tables. 

\noindent{\bf Data acquisition queries.} In this section, we use the term {\em target graph} and {\em join path} interchangeably. For each dataset, we define three queries $Q_1$, $Q_2$ and $Q_3$ of short, medium, and long join paths respectively. In particular, for TPC-H dataset, $Q_1$, $Q_2$ and $Q_3$ are of join path length 2, 3, and 5, while for TPC-E dataset, $Q_1$, $Q_2$ and $Q_3$ are of join path length 3, 5, and 8.  

\noindent{\bf Pricing model \& budget.} We use the entropy-based model for the data marketplace \cite{koutris2015query} to assign the price to data. To simulate the various budget settings, for each acquisition query, we pick the minimum and maximum price of all possible paths between source and target vertices as the lowerbound $LB$ and upperbound $UB$ of the total budget. We define the shopper's budget as $r\times UB$, where $r\in (0,1]$ is the budget ratio. Intuitively, 
the larger $r$ is, the more budget that the shopper can purchase the data. We require that $r\times UB \geq LB$, i.e., the shopper can afford to purchase the instances of at least one target graph in the join graph. 


\noindent{\bf Evaluation metric.} To evaluate the performance of our sampling-based heuristic algorithm, we compare the correlation of the identified data instances by our algorithm and two optimal algorithms, namely the {\em local optimal (LP)} algorithm and the {\em global optimal (GP)} algorithm. Both LP and GP algorithms are the brute-force algorithms that enumerate all paths to find the one of the highest correlation. LP algorithm uses the samples as the input, and the GP algorithm uses the original datasets as the input. For all the algorithms, we measure the real correlation, not the estimated value. 
Let $X_{OPT}$ and $X$ be the correlation of the output by the (LP/GP) optimal approach and our heuristic approach respectively. 
The {\em correlation difference} of is measured as $CD=\frac{X_{OPT} - X}{X_{OPT}}$. Intuitively, the 
smaller correlation difference is, the more accurate our heuristic algorithm is.

\subsection{Scalability}

First, we measure the time performance of our heuristic algorithm with various number of data instances. In Figure \ref{fig:timeperformance-numinstance}, we compare the time performance of our heuristic algorithm against LP and GP on TPC-H data. First, we observe that our heuristic algorithm is significantly more efficient than the two optimal algorithms, especially when $n$ (the number of instances) is large. For example, when $n=8$, our heuristic algorithm can be 2,000 times more efficient than LP, and 20,000 times more efficient than GP. 
We are aware that for $Q_3$, the heuristic algorithm has comparable time performance to LP when $n=5$. This is because for this case, there is only a single I-graph that connects the source and target vertices, which lead to the same search space of our heuristic algorithm and LP. 
Second, we observe that the time performance of the heuristic algorithm is stable when the number of instances varies. Since the number of instances is small, the I-graph discovered by Step 1 (Section \ref{sc:step1}) remains the same. So does the search space of Step 2 (Section \ref{sc:step2}). Therefore, the time performance keeps stable.   
Third, we observe that the time of both LP and GP increase with the growth of $n$. This is not surprising since the number of I-graphs that connect the source and target vertices increase with the growth of $n$. 

Figure \ref{fig:timeperformance-numinstance-e} (a) and (b) show the time performance on TPC-E dataset. Since the two optimal algorithms do no halt within 10 hours on this large dataset, we only show the time performance of our heuristic algorithm in Figure \ref{fig:timeperformance-numinstance-e} (a), and present the I-graph size (i.e., the number of vertices in the graph) for these settings in Figure \ref{fig:timeperformance-numinstance-e} (b). 
An important observation is that the time performance does not increase with the growth of $n$. This is because when we vary the number of instances, the I-graph size indeed fluctuates considerably. However, we do observe that, in general, our heuristic algorithm takes more time when the I-graph is larger. For example, when the I-graph size is 8, it takes 1,010 seconds; while when the I-graph only includes 3 vertices, it can be as fast as 7.3 seconds. 

\begin{figure*}[!htpb]
\centering
\begin{tabular}{ccc}
	\includegraphics[width=0.33\textwidth]{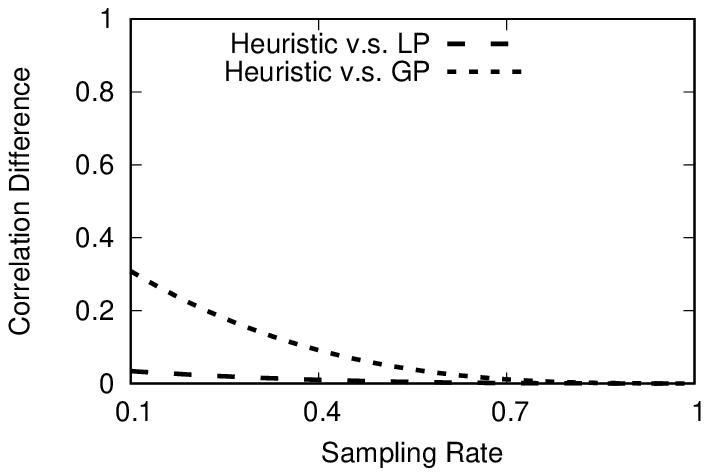}
    &
    \includegraphics[width=0.33\textwidth]{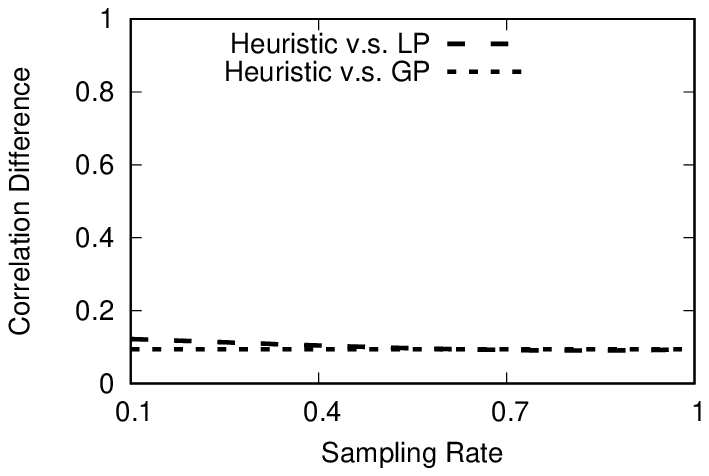}
    &
     \includegraphics[width=0.33\textwidth]{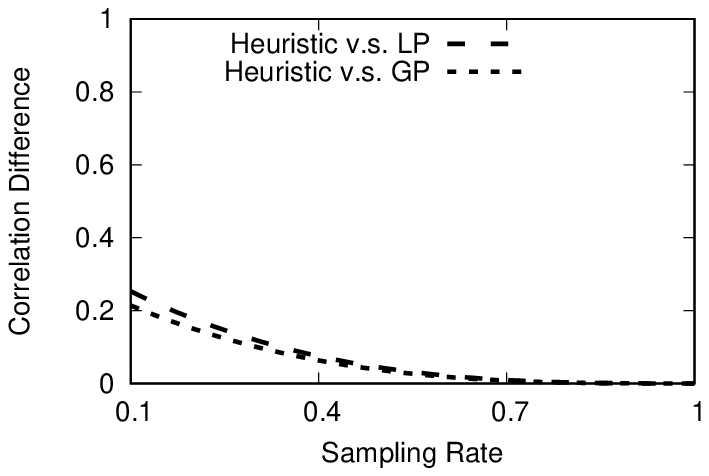}
    \\
    {\small (a) $Q_1$}
    &
    {\small (b) $Q_2$}
    &
    {\small (c) $Q_3$}
\end{tabular}
\vspace{-.1in}
\caption{\small \label{fig:accuracy-sampling} Correlation difference w.r.t. various sampling rate (TPC-H dataset) } 
\vspace{-.1in}
\end{figure*}

\begin{figure*}[!htpb]
\centering
\begin{tabular}{ccc}
	\includegraphics[width=0.33\textwidth]{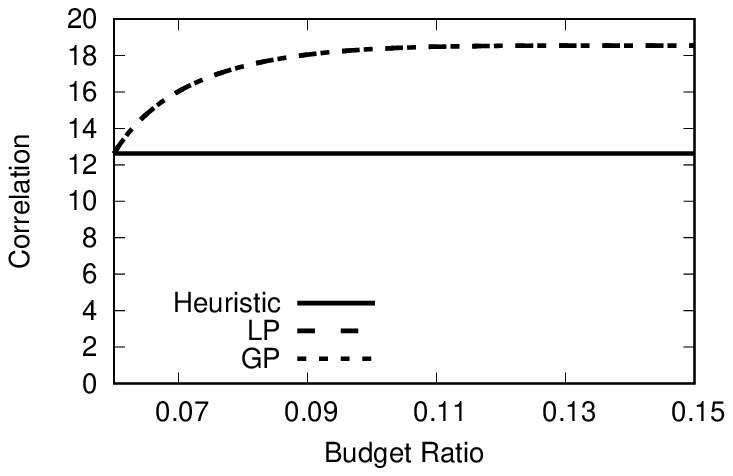}
    &
    \includegraphics[width=0.33\textwidth]{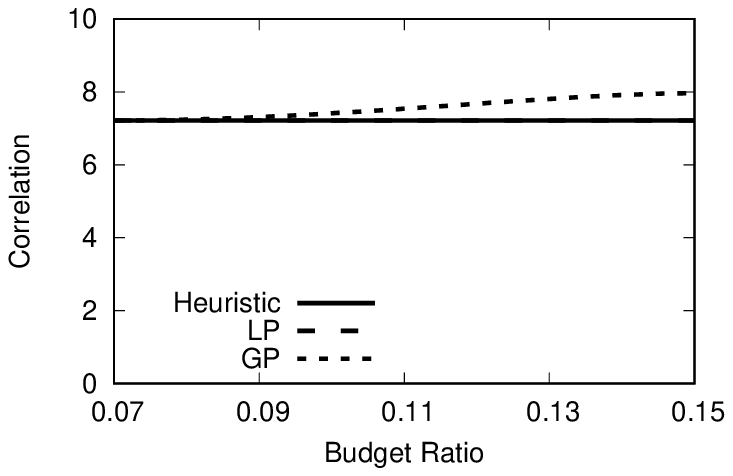}
    &
     \includegraphics[width=0.33\textwidth]{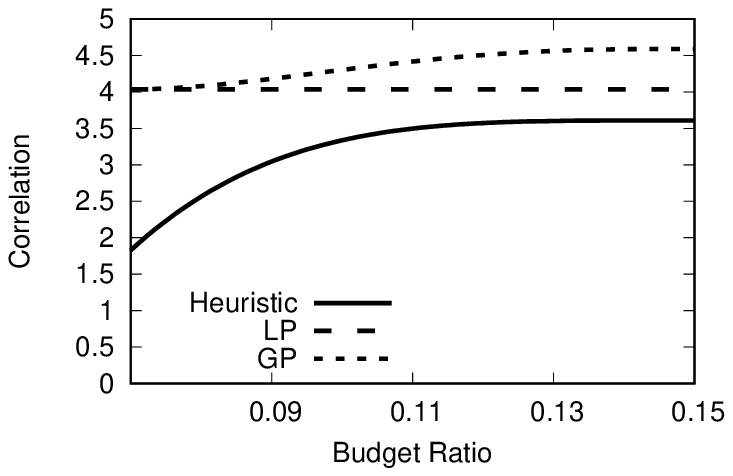}
    \\
    {\small (a) $Q_1$}
    &
    {\small (b) $Q_2$}
    &
    {\small (c) $Q_3$}
\end{tabular}
\vspace{-.1in}
\caption{\small \label{fig:accuracy-budget} Correlation w.r.t. various budget ratio (TPC-H dataset)} 
\vspace{-.1in}
\end{figure*}

\begin{figure*}[!hbt]
	\centering
	\begin{tabular}{ccc}
	\includegraphics[width=0.33\textwidth]{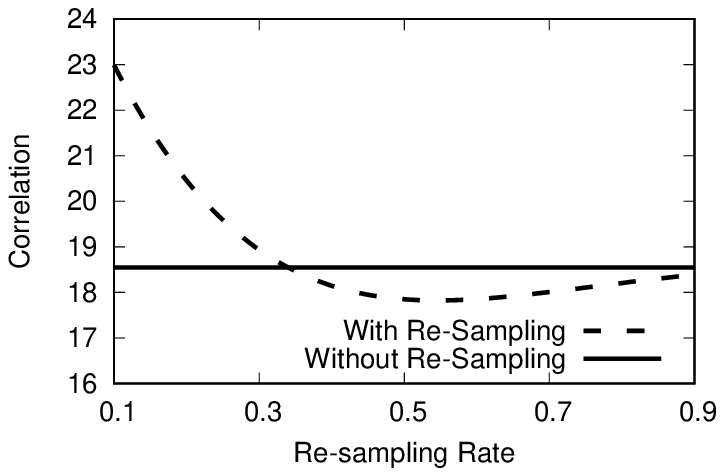}
    &
	\includegraphics[width=0.33\textwidth]{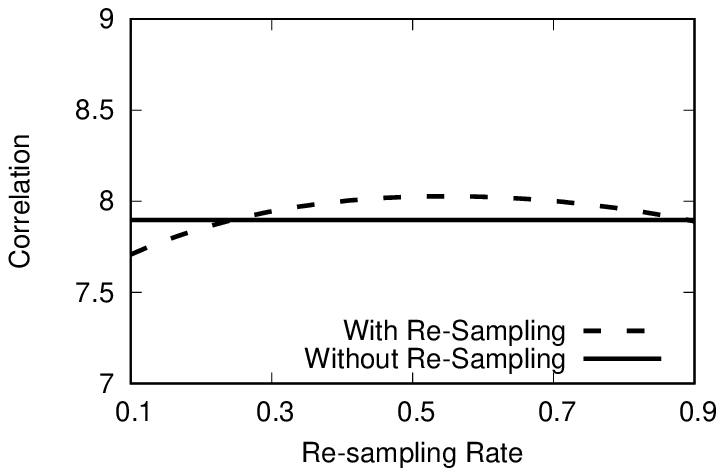}
    &
	\includegraphics[width=0.33\textwidth]{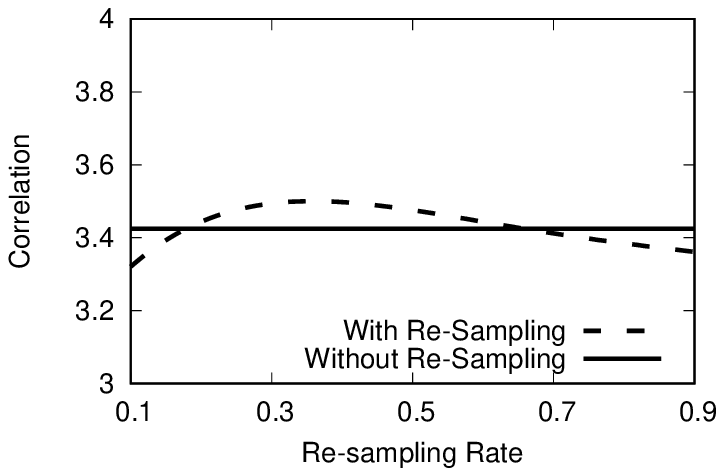}
    \\
    {\small (a) $Q_1$}
    &
    {\small (b) $Q_2$}
    &
    {\small (c) $Q_3$}
\end{tabular}
\vspace{-.1in}
 \caption{\small \label{fig:accuracy-nosampling} Correlation with and without re-sampling w.r.t. various re-sampling rates (TPC-H dataset) } 
 \vspace{-.1in}
\end{figure*}

We also measure the time performance of our heuristic algorithm with regard to various budget ratios. 
Figure \ref{fig:timeperformance-numinstance-e} (c) shows the time performance of our heuristic algorithm with various budget ratios on TPC-E dataset. We vary the budget ratio from 0.04 to 0.12, where 0.04 is the minimum budget ratio that can find at least one solution for either of the three queries $Q_1$, $Q_2$ and $Q_3$, while 0.12 is the highest budget ratio that can find at least one solution for all the three queries. With a small budget ratio, \system may not be able to find an affordable acquisition solution for some queries. For example, there is no acquisition solution for $Q_3$ when the budget ratio is under 0.06 (noted as N/A in Figure \ref{fig:timeperformance-numinstance-e} (c)). 
From the results, first, we observe that \system takes more time when the budget ratio grows.  However, the time performance may keep stable when the budget ratio is sufficiently large. For example, when the budget ratio is higher than 0.08, the time performance of $Q_1$ keeps unchanged. This is because with budget ratio 0.08, every target graph sample in Algorithm \ref{alg:mcmc}  becomes affordable. For this case, \system  calculates the correlation of all the samples to find the one with the highest correlation. The number of samples keeps the same even if we increase the budget.  Second, \system takes more time to process the acquisition requests of  longer join paths. This is straightforward since it takes more time to join the instances in the longer path, and estimate the correlation and quality.
We also measure the time performance on TPC-H dataset, and have similar observation as TPC-E dataset. Due to the space limit, we present the detailed results in our full paper \cite{cost2018li}.


\subsection{Correlation}

\nop{
\begin{figure}[!hbt]
\vspace{-0.1in}
	\centering
	\begin{tabular}{@{}c@{}c@{}}
   &
	\\
	{\scriptsize(a) TPC-H dataset}
	&
	{\scriptsize(b) TPC-E dataset}
	\end{tabular}
 \caption{\small \label{fig:correlation-price} The effect of pricing functions on correlation} 
\end{figure}
}

First, we measure the difference of the correlation of source and target attribute sets in the datasets returned by our heuristic algorithm and the two optimal algorithms with regard to various sampling rates. Intuitively, we want the correlation difference to be close to 0. The results are shown in Figure \ref{fig:accuracy-sampling}. First, we notice that the correlation difference is very small. In all cases, it never exceeds 0.31. Recall that in Figure \ref{fig:timeperformance-numinstance}, our heuristic algorithm can be 2,000 times more efficient than the optimal methods. This demonstrates that our method can efficiently find the datasets of correlation that is comparable to the optimal results. 
Second, we observe that the correlation difference decreases when the sampling rate grows. This is straightforward as more samples lead to more accurate correlation estimation. 

Second, we measure the correlation between source and target attributes in the data acquisition result returned by our heuristic algorithm, as well as in the two optimal algorithms, on TPC-H dataset. The comparison results are displayed in Figure \ref{fig:accuracy-budget}. First, we notice that the correlation by our heuristic algorithm is close to that of both optimal algorithms. In all cases, the difference is at most 5.9 (the maximum correlation is around 19). 
Second, with the increase of the budget ratio, the correlation in the results returned by all the three algorithms gradually rise (i.e., the correlation gets stronger). This is straightforward, as higher budget can afford to purchase more data with better utility.
We do not show the correlation measurement results on TPC-E dataset, due to the long execution time of the GP algorithm. 

We also measure the impacts of re-sampling on the correlation by changing the re-sampling rate and measuring the correlation of the acquisition result by our heuristic algorithm with and without re-sampling. The result is presented in Figure \ref{fig:accuracy-nosampling}. We observe that the correlation with re-sampling oscillates around the correlation without re-sampling. The difference gradually reaches 0 with the growth of the re-sampling rate. Overall, the estimated correlation with re-sampling is accurate. The difference with the estimated correlation without re-sampling never exceeds 4.5.

\subsection{Data Acquisition with \system vs. without \system}
We compare the data acquisition results (correlation, data quality, join informativeness, and price) by \system with direct purchase from the data marketplace. We use GP algorithm to find the data acquisition results on the data marketplace. The comparison result is displayed in Table \ref{table:results}. First, we observe a large overlap between the data acquisition results returned by \system and GP. For instance, for $Q_3$, the data acquisition results returned by \system is \{orders(totalprice, custkey), customer(custkey, H), supplier(H, nationkey), nation(nationkey, regionkey), region\\(regionkey, rname)\}, where H is one of the fake join attribute that is added to the data, while the results returned by GP is \{orders(totalprice, custkey), customer(custkey, nationkey, H), supplier(nationkey, H), nation(nationkey, regionkey), region(regionkey, rname)\}. Due to the space limit, we omit the target graph discovered by both approaches in Table \ref{table:results}.  
Second, we observe that the correlation of the data acquisition results returned by \system is comparable to that of GP. It can be as high as 90\% of the optimal result. 
Third, the join informativeness of the data acquisition by both \system and GP is also close, which demonstrates the superiority of our correlated re-sampling method. 
The price of the data acquisition results returned by \system is always lower than that of GP. For example, it is 42\% lower than the price of  $Q_2$ on the data marketplace. This shows that \system is able to find the data of high utility (correlation) at a lower price. 
We acknowledge that in some cases, the quality of the data acquisition results returned by \system (e.g., $Q_1$) is significantly lower than that of GP, due to the error introduced by the sampling-based estimation. However, in most cases, the accuracy is still satisfying ($Q_2$ and $Q_3$).

\nop{
\begin{table*}[h!]
\begin{center}
\begin{tabular}{|c|c|c|c|c|c|c|}\hline
   Approach & Query & Join path & Correlation & Quality & Join Informativeness & Price \\\hline 
  \multirow{3}{*}{Purchase from data marketplace} & $Q_1$ & customer(cacctbal,G), orders(G,clerk) & 18.18  & 0.3819 & 0.8268 & 63.12 \\\cline{2-7} 
  & $Q_2$ & nation(nname,nationkey) supplier(nationkey,F) lineitem(F,suppkey) partsupp(suppkey,availqty) & 7.965 & 0.4244 & 1.040 & 103.2 \\\cline{2-7} 
  & $Q_3$ & orders(totalprice,custkey) customer(custkey,nationkey,H) supplier(nationkey,H) nation(nationkey,regionkey) region(regionkey,rname)  & 4.592 & 0.1026 & 2.104 & 106.4 \\\hline 
 \multirow{3}{*}{With \system} & $Q_1$ & customer(cacctbal,custkey), orders(custkey,clerk) & 12.51  & 0.07688 & 0.8906 & 60.65\\\cline{2-7} 
  & $Q_2$ & nation(nname,nationkey) supplier(nationkey,E) partsupp(E,availqty) & 7.216 & 0.3968 & 1.248 & 59.78\\\cline{2-7}
  & $Q_3$ & orders(totalprice,custkey) customer(custkey,H) supplier(H,nationkey) nation(nationkey,regionkey) region(regionkey,rname) & 3.609 &  0.08632 & 2.063 & 100.5 \\\hline                          
 \end{tabular}
\caption{\label{table:results} Comparison between data acquisition with \system  and purchase from the data marketplace directly (TPC-H dataset, budget ratio=0.13) }
\end{center}
\end{table*}
\Wendy{Compare heuristic with GP algorithm.Results shown in Table \ref{table:results}.}
}

\vspace{-0.1in}
\section{Related Work}
\label{sc:related}

The concept of {\em data market} is firstly formally defined in \cite{balazinska2011data}. 
Kanza et al. \cite{kanza2015online} envision a geo-social data marketplace that facilitates the generation and selling of high-quality spatio-temporal data from people. 
Koutris et al. \cite{koutris2015query} propose a query-based data pricing model for the market, considering the complicated queries that involve conjunctive sub-queries. 
The history-aware pricing model \cite{upadhyaya2016price} avoids the buyer from getting charged repeatedly for the same data returned by different queries.
Ren et al. \cite{ren2016joint} focus on the joint problem of data purchasing and data placement in a cloud data market. 
None of them studies the correlation-driven purchase on the data marketplace.
The key to the data market model is the pricing model. Balazinska et al. \cite{balazinska2011data} propose the data pricing model in which the buyers are charged based on the queries. A fundamental requirement named {\em arbitrage-free requirement} is identified in the pricing function, in order to prevent the buyers from intentionally reducing the charge by decomposing a complex query into multiple simple ones. 
QueryMarket \cite{koutris2012querymarket} demonstrates the superiority of such a query-based data pricing model in real-world applications.
Koutris et al. \cite{koutris2015query} prove that it is NP-complete  to determine the arbitrage-free price of certain conjunctive selection queries for a given database instance.
Lin et al. \cite{lin2014arbitrage} study the arbitrage-free requirement theoretically and discover the sufficient conditions to enforce it in two models. The first model is {\em instance-independent}, in which the pricing function depends on the query only, and is irrelevant to the database instance. The other one is named the {\em delayed model} in which the price is determined by both the query and the result. Based on the findings of \cite{lin2014arbitrage}, Deep et al. \cite{deep2016design} prove that any monotone, submodular and subadditive pricing function satisfies the arbitrage-free query pricing requirement. A survey of pricing models can be found in \cite{muschalle2012pricing}.

A relevant line of work is to explore the databases that contain complex database schemas via joins. Intuitively, given a database with schema graph,  in which two tables are specified as the source and destination tables, the problem is to find a valid join path between the source and destination. 
To solve this problem, Procopiuc et al. \cite{procopiuc2008database} compute a probability for each join path that connects a given source table to a destination table. 
To speed up the exploration over the complex schema graph, Yang et al. \cite{yang2011summary}  propose a new approach that 
summarizes the contents of a relational database as 
a {\em summary graph}. The most relevant tables and join joins can be computed from the summary graph. It defines the importance of each table
in the database as its stable state value in a random walk over
the schema graph, where the transition probability is defined based on the
entropies of table attributes. 
Zhang et al. \cite{zhang2013reverse} take the reverse-engineering approach. In particular, given a database $D$ with schema graph $G$ and an output table $Out$, the problem is to compute a join query $Q$ that generates $Out$ from $D$. Unlike these work that mainly focus on the join informativeness of data instances, we take multiple factors, including join informativeness, data quality, and price, into consideration for the data marketplace.

Regarding the related work on subgraph search, the work on centerpiece subgraphs \cite{tong2007fast,tong2006center}, and on entity-relationship subgraphs \cite{kasneci2009ming} partially share with our goal: for a given set of query nodes and a budget
constraint, output a connecting subgraph that meets the budget and
maximizes some goodness measure. These works mainly focus on mining the patterns of some specific subgraphs, while we focus on finding the optimal subgraph with regards to the given constraints on informativeness, quality, and price. 





\nop{
Pipino et al. \cite{pipino2002data} propose a list of principles of data quality measurement. One important evaluation metric is data consistency with regard to the {\em integrity constraints} like {\em functional dependencies (FDs)}. Chomicki et al. \cite{chomicki2005minimal} leverage FDs to detect the existence of data inconsistency and fix the issue by deleting a small fraction of tuples. 
Bohannon et al. \cite{bohannon2005cost} remove data inconsistency (i.e., the records that do not follow FDs) by applying changes to the data values. They prove that it is a NP-complete problem to find the minimum-cost repair, and thus design an efficient greedy algorithm. 
Beskales et al. \cite{beskales2010sampling} revise the outdated FDs to make the updated FDs consistent with the data. 
Chiang et al. \cite{chiang2011unified} combine the ideas in \cite{bohannon2005cost, beskales2010sampling} by proposing a unified model in which both the data values and FDs can be adjusted to reach consistency. 
}

\section{Conclusion}
\label{sc:conclusion}
In this paper, we study the data acquisition problem for correlation analysis in the data marketplace framework. We consider quality, join informativeness, and price issues of data acquisition, and model the data acquisition problem as a graph search problem. We  prove that the graph search problem is NP-hard, and design a heuristic algorithm based on Markov chain Monte Carlo (MCMC). Our experiment results demonstrate the performance of our algorithm.

For the future, there are quite a few interesting research directions to explore. For example, instead of the best acquisition scheme, \system may recommend a number of acquisition options of the top-k scores to the data buyer, where the scores can be defined as a combination of correlation, data quality, join informativeness, and price. This raises the issues of how to define a fair score function, as well as the design of efficient top-k search algorithm when the score function is not be monotone. Additional issues that are worth to explore include how to deal with {\em heterogeneous} data in the marketplace, and how to deal with fast-evolving data in the data marketplace. 
Furthermore, an important issue is the privacy of the data marketplaces. In particular, when \system is not fully trusted, how can the data shopper exchange the information of the source instances with \system in a privacy-preserving way, and find correlated instances in the data marketplace? How to verify if the data in the marketplace is genuine (i.e., it is not fabricated)?

\bibliographystyle{abbrv}
{
\bibliography{bib_short}  

\begin{thebibliography}{10}

\bibitem{bdex}
Bdex.
\newblock http://www.bigdataexchange.com/.

\bibitem{gwarehouse}
Google bigquery data warehouse.
\newblock https://cloud.google.com/bigquery/.

\bibitem{azure}
Microsoft azure data marketplace.
\newblock
  https://azuremarketplace.microsoft.com/en-us/marketplace/?source=datamarket.

\bibitem{IDC}
Idc's worldwide semiannual big data and analytics spending guide taxonomy.
\newblock \url{http://www.informationweek.com/big-data/}, 2016.

\bibitem{arenas1999consistent}
M.~Arenas, L.~Bertossi, and J.~Chomicki.
\newblock Consistent query answers in inconsistent databases.
\newblock In {\em Proceedings of the ACM Symposium on Principles of Database
  Systems}, pages 68--79, 1999.

\bibitem{balazinska2011data}
M.~Balazinska, B.~Howe, and D.~Suciu.
\newblock Data markets in the cloud: An opportunity for the database community.
\newblock {\em Proceedings of the VLDB Endowment}, 4(12):1482--1485, 2011.

\bibitem{chiang2008discovering}
F.~Chiang and R.~J. Miller.
\newblock Discovering data quality rules.
\newblock {\em Proceedings of the VLDB Endowment}, 1(1):1166--1177, 2008.

\bibitem{deep2016design}
S.~Deep and P.~Koutris.
\newblock The design of arbitrage-free data pricing schemes.
\newblock In {\em International Conference on Database Theory}, 2017.

\bibitem{fan2008conditional}
W.~Fan, F.~Geerts, X.~Jia, and A.~Kementsietsidis.
\newblock Conditional functional dependencies for capturing data
  inconsistencies.
\newblock {\em ACM Transactions on Database Systems}, 33(2):6, 2008.

\bibitem{Gubichev:2010:FAE:1871437.1871503}
A.~Gubichev, S.~Bedathur, S.~Seufert, and G.~Weikum.
\newblock Fast and accurate estimation of shortest paths in large graphs.
\newblock In {\em Proceedings of ACM International Conference on Information
  and Knowledge Management}, pages 499--508, 2010.

\bibitem{hague2013market}
P.~N. Hague, N.~Hague, and C.-A. Morgan.
\newblock {\em Market research in practice: How to get greater insight from
  your market}.
\newblock Kogan Page Publishers, 2013.

\bibitem{huhtala1999tane}
Y.~Huhtala et~al.
\newblock Tane: An efficient algorithm for discovering functional and
  approximate dependencies.
\newblock {\em The Computer Journal}, 1999.

\bibitem{kanza2015online}
Y.~Kanza and H.~Samet.
\newblock An online marketplace for geosocial data.
\newblock In {\em Proceedings of the ACM International Conference on Advances
  in Geographic Information Systems}, page~10, 2015.

\bibitem{kasneci2009ming}
G.~Kasneci, S.~Elbassuoni, and G.~Weikum.
\newblock Ming: mining informative entity relationship subgraphs.
\newblock In {\em Proceedings of the ACM Conference on Information and
  Knowledge Management}, pages 1653--1656, 2009.

\bibitem{koutris2012querymarket}
P.~Koutris, P.~Upadhyaya, M.~Balazinska, B.~Howe, and D.~Suciu.
\newblock Querymarket demonstration: Pricing for online data markets.
\newblock {\em Proc. of the VLDB Endowment}, 5(12):1962--1965, 2012.

\bibitem{koutris2015query}
P.~Koutris, P.~Upadhyaya, M.~Balazinska, B.~Howe, and D.~Suciu.
\newblock Query-based data pricing.
\newblock {\em Journal of the ACM}, 62(5):43, 2015.

\bibitem{cost2018li}
Y.~Li, H.~Sun, B.~Dong, and W.~H. Wang.
\newblock Cost-efficient data acquisition on online data marketplaces for
  correlation analysis (full version).
\newblock Technical report, 2018.
\newblock Available at
  \url{https://msuweb.montclair.edu/~dongb/publications/dacron-full.pdf}.

\bibitem{lin2014arbitrage}
B.-R. Lin and D.~Kifer.
\newblock On arbitrage-free pricing for general data queries.
\newblock {\em Proceedings of the VLDB Endowment}, 7(9):757--768, 2014.

\bibitem{muschalle2012pricing}
A.~Muschalle, F.~Stahl, A.~L{\"o}ser, and G.~Vossen.
\newblock Pricing approaches for data markets.
\newblock In {\em International Workshop on Business Intelligence for the
  Real-time Enterprise}, pages 129--144, 2012.

\bibitem{nguyen2014detecting}
H.~V. Nguyen, E.~M{\"u}ller, P.~Andritsos, and K.~B{\"o}hm.
\newblock Detecting correlated columns in relational databases with mixed data
  types.
\newblock In {\em Proceedings of the International Conference on Scientific and
  Statistical Database Management}, page~30, 2014.

\bibitem{procopiuc2008database}
C.~M. Procopiuc and D.~Srivastava.
\newblock Database exploration using join paths.
\newblock In {\em IEEE International Conference on Data Engineering}, pages
  1331--1333. IEEE, 2008.

\bibitem{qian2012sample}
L.~Qian, M.~J. Cafarella, and H.~Jagadish.
\newblock Sample-driven schema mapping.
\newblock In {\em Proceedings of the ACM International Conference on Management
  of Data}, pages 73--84. ACM, 2012.

\bibitem{rahm2000data}
E.~Rahm and H.~H. Do.
\newblock Data cleaning: Problems and current approaches.
\newblock {\em IEEE Data Engineering Bulletin}, 23(4):3--13, 2000.

\bibitem{ren2016joint}
X.~Ren et~al.
\newblock Joint data purchasing and data placement in a geo-distributed data
  market.
\newblock {\em arXiv preprint arXiv:1604.02533}, 2016.

\bibitem{song2010efficient}
S.~Song and L.~Chen.
\newblock Efficient set-correlation operator inside databases.
\newblock In {\em Proceedings of the International Conference on Information
  and Knowledge Management}, pages 139--148. ACM, 2010.

\bibitem{tong2006center}
H.~Tong and C.~Faloutsos.
\newblock Center-piece subgraphs: problem definition and fast solutions.
\newblock In {\em Proceedings of the ACM International Conference on Knowledge
  Discovery and Data Mining}, pages 404--413, 2006.

\bibitem{tong2007fast}
H.~Tong, C.~Faloutsos, and Y.~Koren.
\newblock Fast direction-aware proximity for graph mining.
\newblock In {\em Proceedings of the 13th ACM SIGKDD international conference
  on Knowledge discovery and data mining}, pages 747--756, 2007.

\bibitem{upadhyaya2016price}
P.~Upadhyaya et~al.
\newblock Price-optimal querying with data apis.
\newblock {\em Proceedings of the VLDB Endowment}, 2016.

\bibitem{vazirani2013approximation}
V.~V. Vazirani.
\newblock {\em Approximation algorithms}.
\newblock Springer Science \& Business Media, 2013.

\bibitem{vengerov2015join}
D.~Vengerov, A.~C. Menck, M.~Zait, and S.~P. Chakkappen.
\newblock Join size estimation subject to filter conditions.
\newblock {\em Proceedings of the VLDB Endowment}, 8(12):1530--1541, 2015.

\bibitem{yang2009summarizing}
X.~Yang, C.~M. Procopiuc, and D.~Srivastava.
\newblock Summarizing relational databases.
\newblock {\em Proceedings of the VLDB Endowment}, 2(1):634--645, 2009.

\bibitem{DBLP:journals/pvldb/YangPS11a}
X.~Yang, C.~M. Procopiuc, and D.~Srivastava.
\newblock Summary graphs for relational database schemas.
\newblock {\em {Proceedings of the VLDB Endowment}}, 4(11):899--910, 2011.

\bibitem{yang2011summary}
X.~Yang, C.~M. Procopiuc, and D.~Srivastava.
\newblock Summary graphs for relational database schemas.
\newblock {\em Proceedings of the VLDB Endowment}, 2011.

\bibitem{zhang2013reverse}
M.~Zhang, H.~Elmeleegy, C.~M. Procopiuc, and D.~Srivastava.
\newblock Reverse engineering complex join queries.
\newblock In {\em Proceedings of the ACM International Conference on Management
  of Data}, pages 809--820, 2013.

\end{thebibliography}
}

\end{document}